\documentclass[11pt,a4paper]{article}

\usepackage[english]{babel}

\usepackage{natbib}
\usepackage{amsmath}
\usepackage{amsfonts}
\usepackage{amssymb}
\usepackage{float}
\usepackage{microtype}
\usepackage{graphicx}
\usepackage{fullpage}

\usepackage{amsmath}
\usepackage{amsfonts}
\usepackage{amsthm}
\usepackage{graphicx}

\usepackage{algorithm}
\usepackage{algorithmic}
\usepackage{natbib}
\usepackage{float}
\usepackage{url}
\usepackage{microtype}

\usepackage{epstopdf}

\usepackage{parskip}

\usepackage[font=small,format=plain,labelfont=bf,up,textfont=it,up]{caption}

\usepackage{bm}
\newcommand{\bmm}[1]{\mathbf{#1}}

\title{The use of systems of stochastic PDEs as priors for multivariate models with discrete structures\\
}

\author{Erlend Aune and Daniel P.~Simpson \\
Department of Mathematical Sciences, NTNU, Norway}

\begin{document}

\maketitle

\abstract 

A challenge in multivariate problems with discrete structures is the inclusion of prior information that may differ in each separate structure. A particular example of this is seismic amplitude versus angle (AVA) inversion to elastic parameters, where the discrete structures are geologic layers. Recently, the use of systems of linear stocastic partial differential equations (SPDEs) have become a popular tool for specifying priors in latent Gaussian models. This approach allows for flexible incorporation of nonstationarity and anisotropy in the prior model. Another advantage is that the prior field is Markovian and therefore the precision matrix is very sparse, introducing huge computational and memory benefits. We present a novel approach for parametrising correlations that differ in the different discrete structures, and additionally a geodesic blending approach for quantifying fuzziness of interfaces between the structures.\vspace{18pt} 


\section{Introduction}

In spatial statistics, the need for specifying different behaviour in different regions in space is crucial for making a good prior model. The litterature is abundant with methodologies for this. In the multivariate setting, this generalises to having different correlations between the fields in different regions and different cross-differentiability properties.

A particular model problem where this is important is the seismic AVA inversion problem, which well studied in the geophysical litterature. There are several incarnations of this problem with varying degrees of complexity. In this article, our primary example is the inversion problem studied in \cite{bul_omr_linavo,bul_et_al_rapidavo,rab_urs_nonlinsampling}, using the wavefield propagation approximations in \cite{aki1980quantitative}, which results in linear systems of equations to solve. Variants and extensions of these equations are found in \cite{sto_urs_ref}, including nonlinear approximations that may yield better inversion results in some situations. We exemplify our contributions using this example explicitly throughout the article.

The model we adopt in this text is the same as in \cite{bul_omr_linavo}, which is essentially
\begin{align}
	d(\bmm{s}) = w \star r_{PP}(\bmm{m}) (\bmm{s}) + \epsilon, 
\end{align}
where $\star$ denotes convolution in time, $w$ is an approximation to the source wavelet -- i.e. the shape of the wave traveling through the subsurface, and $r_{PP}(\bmm{m})$ denotes a reflectivity operator. The reflectivity operator takes relative differences in elastic parameters to reflection coefficients for the wave. We adopt the following elastic parameters,
\begin{align}
	m_1=\frac{\triangle v_P}{\overline{v}_P}, \quad m_2=\frac{\triangle v_S}{\overline{v}_S}, \quad m_3=\frac{ \triangle \rho}{ \overline{\rho}}.
\end{align}
I.e. $m_1,m_2,m_3$ denotes the relative difference of $P$-velocity (pressure wave velocity), $S$-velocity (shear wave velocity) and density respectively, and the reflectivity operator is defined by
\begin{align}
	r_{PP}(\theta)=\frac{m_1}{2}(1+\tan^2 \theta)   -4 m_2 \gamma^2 \sin^2 \theta +\frac{m_3}{2}(1-4 \gamma^2 \sin^2 \theta),
\end{align}
there $\theta$ is the reflection angle and $\gamma^2$ denotes a background $(v_S/v_P)$-ratio. Rewriting this in matrix notation yields
\begin{align}
	\bmm{d} =\bmm{WAm} + \bm{\epsilon},
\end{align}
where $\bmm{d}$ are observations, $\bmm{W}$ is the discretized wavelet operator, $\bmm{A}$, the discretized reflectivity operator, $\bmm{m}$ the elastic parameters and $\bm{\epsilon} \sim \mathcal{N}(\bm{0},\sigma^2 \bmm{I})$ an error term which is often assumed to be normally distributed.

In this text, we will explore a novel method for designing a good prior for $\bmm{m}$ using linear systems of stochastic partial differential equations. We emphasize, however, that while the approach developed here is designed with seismic AVA inversion in mind, it is very flexible and can be adopted in any setting where we have multivariate fields with separate regions where we would like to incorporate prior information.

All the figures that appear in this text have comparative scales, so that the colour schemes have the same min-max values in each individual figure. Hence, the figures makes sense, without cluttering them with additional colourbars.

\section{Prior specification}

The choice of prior in the inversion problem is of great importance when it comes to the performance of the inversion. It is vital to choose a ``good'' prior to emphasise the properties of $\bmm{m}$ that we know it has. For us, $\bmm{m}$ will denote the parameters of interest, and it depends on position. We construct the prior by combining heuristics and expert knowledge of the spatial model. For a Gaussian prior model, the standard way of specifying the prior model is through the covariance function, which is often assumed to be stationary (see, e.g. \cite{bul_omr_linavo}). A stationary covariance function is defined by a correlation function that defines how much a point is correlated with its neighbours and a marginal variance parameter, $\varrho^2$ through
\begin{align}
	\varrho^2 c(\| \bmm{x} - \bmm{y} \|_\bmm{A})=\text{Cov}(\bmm{x},\bmm{y}), \label{eq_covfung_geos}
\end{align}
where $\bmm{A}$ is a positive definite matrix that defines the non-changing anisotropy of the field. In the Gaussian case, this defines a strictly stationary process if the mean is constant. There is a list of widely used covariance functions in \cite{cressie_spatdata}. We will throughout this text assume that the prior is from the Gaussian family. This family is defined by having density
\begin{align}
	p(\bmm{x}|\bmm{Q},\bm{\mu}_x) = (2 \pi)^{-n/2} \det (\bmm{Q})^{1/2} \exp \left( \frac{1}{2} (\bmm{x}-\bm{\mu}_x)^T \bmm{Q} (\bmm{x} - \bm{\mu}_x) \right), \label{eq_gauss_simple}
\end{align}
where $\bmm{Q}=\bm{\Sigma}^{-1}$ is the precision matrix -- the inverse of the covariance matrix $\bm{\Sigma}$ -- and $\bm{\mu}_x$ is the expectation, $\mathbb{E}(\bmm{x} | \bm{\mu}_x)$.

Moreover, the fields $m_1,m_2,m_3$ are assumed correlated with correlations specified by well data and/or other local knowledge. In the discretized domain, this allows for the following decomposition of the total covariance matrix
\begin{align}
	\bm{\Sigma}_m = \bm{\Sigma}_\text{space} \otimes \bm{\Sigma}_0 \label{eq_prior_cov_spec}
\end{align}
where $\bm{\Sigma}_\text{space}$ denotes the spatial covariance matrix, typically defined through a covariance function, and $\bm{\Sigma}_0$ the correlations between the elastic parameters. Since seismic observations typically are on a regular grid, either in 2-D or 3-D, it possible to let $\bm{\Sigma}_\text{space}$ be circulant by extending the grid by as many points as is needed to get the correlation below a threshold -- typically $0.1$ or $0.05$. This allows us to use fast Fourier transforms for computing quantities of interest related to the covariance matrix. This, together with the Kronecker structure of $\bm{\Sigma}_m$ allows for fast computations. See \cite{bul_et_al_rapidavo,rue_gmrf,gray_circ_toep} for details. This approach also has very low memory requirements; since $\bm{\Sigma}_\text{space}$ is  circulant it may be stored using only one vector. Hence storage is $\mathcal{O}(n)$ and computations (of any kind) are at most $\mathcal{O}(n \log n)$, where $n$ is the number of nodes in the extended lattice.

\subsection{SPDE formulation}
While this decomposition is sensible, it is also very inflexible and requires stationarity for low storage requirements. Another way of pursuing good prior models with fast computations and low memory requirements is through the use of elliptic (pseudo) differential operators (\cite{ruzhansky2009pseudo}, part 2 is an accessible source). The theory of pseudo differential operators is closely related to Weyl transforms and short-time Fourier transforms or Gabor transforms (\cite{feichtinger2008pseudo}) and usual spectral considerations is seismology apply. In this approach, it is the sparsity of the resulting precision matrices that makes storage and computation manageable. Recently, \cite{lindgren2010explicit}, studied how to apply such operators in a statistical setting. They studied a Riesz-Bessel operator, $(-\triangle + \kappa)^{\alpha/2}$ and its relation to computation and Mat\'{e}rn covariance models \citep{matern1986spatial,whittle1963stochastic}. The main lessons are firstly, if
\begin{align}
	M_{\kappa,\alpha} x(s) :=(\kappa^2 - \triangle)^{\alpha/2} x(s) = \mathcal{W}(s), \label{eq_matern_stat}
\end{align}
where $\mathcal{W}$ is spatial Gaussian white noise, then $x(s)$ has Mat\'{e}rn type covariance function, i.e.,
\begin{align}
	\rho(r)  & = \frac{\varrho^2}{\Gamma(\alpha-d/2) 2^{\alpha-d/2-1}} \left(\kappa r \right)^{\alpha-d/2} K_{\alpha-d/2}(\kappa \, r), \\
	\varrho^2 & = \frac{\Gamma(\alpha-d/2)}{\Gamma(\alpha) (4 \pi)^{d/2} \kappa^{2(\alpha-d/2)} },
\end{align} 
where $K_s$ is the modified Bessel function of the first kind. Secondly, fast computations through finite element methods or other discretisations of the differential operator in \eqref{eq_matern_stat} are available through the induced Markov properties of the discretisation matrix, $\bmm{Q}_\text{space}$. That essentially means that $\bmm{Q}_m=\bmm{Q}_\text{space} \otimes \bmm{Q}_0$ is (very) sparse and with a structure ameanable to Cholesky factorisation. An alternative requirement is that we can construct the matrix vector product $\bmm{Q}_m \bmm{v}$ and $\det(\bmm{Q}_m)$ relatively quickly through some iterative or direct procedure, see \cite{simpson_phd,aune_sampling_cgm,proper_logdet}

When addressing the ``stationarity'' of the field defined by \eqref{eq_matern_stat}, it is only stationary in the sense of \eqref{eq_covfung_geos} if it is defined on the whole of $\mathbb{R}^k$, where $k=2,3$ in our case --  alternatively when the corresponding operator is defined on a manifold without boundary. In our case the domain on which \eqref{eq_covfung_geos} is defined is merely a subset, namely a square or box in $\mathbb{R}^2$ or $\mathbb{R}^3$. Hence boundary effects resulting from boundary conditions may destroy its direct interpretability in terms of this equation. It is, of course, possible to specify boundary conditions in such a way that you retain the property in \eqref{eq_covfung_geos}, but usually there are more natural physical boundary conditions that in our opinion improves upon the specification through SPDEs compared to the model defined by covariance matrices through stationary covariance functions also in the stationary case.

\begin{figure}
	\caption{Realisations from stationary model given by \eqref{eq_matern_stat} (left) and non-stationary model given by \eqref{eq_matern_nonstat} (right). The non-stationary model has a curved interface, and the field below the interface has anisotropy directed along the curve, while above the interface it is almost isotropic.} \label{fig_univar_stat_nonstat}
	\begin{center}
	\includegraphics[width=0.45\textwidth]{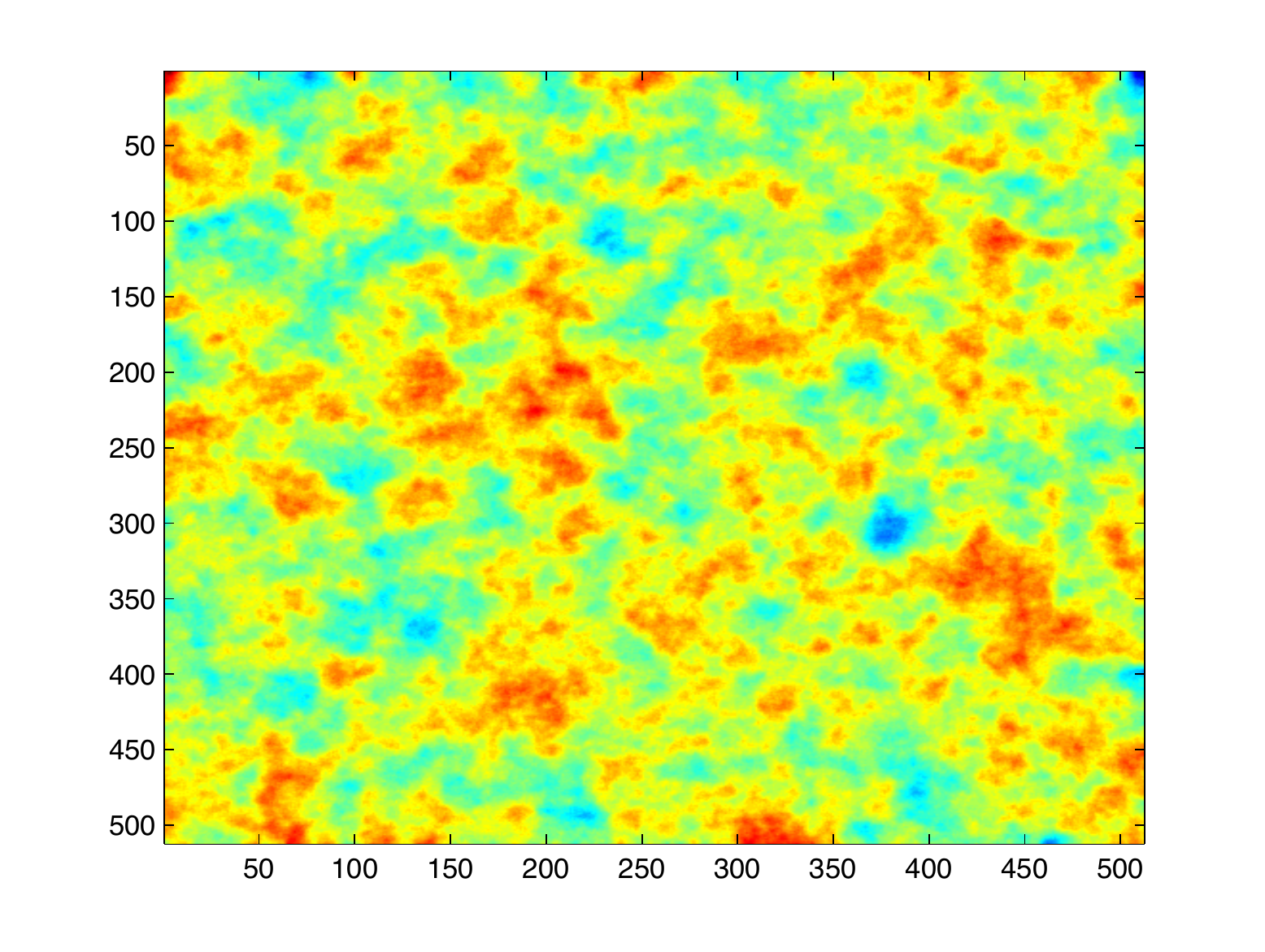} \includegraphics[width=0.45\textwidth]{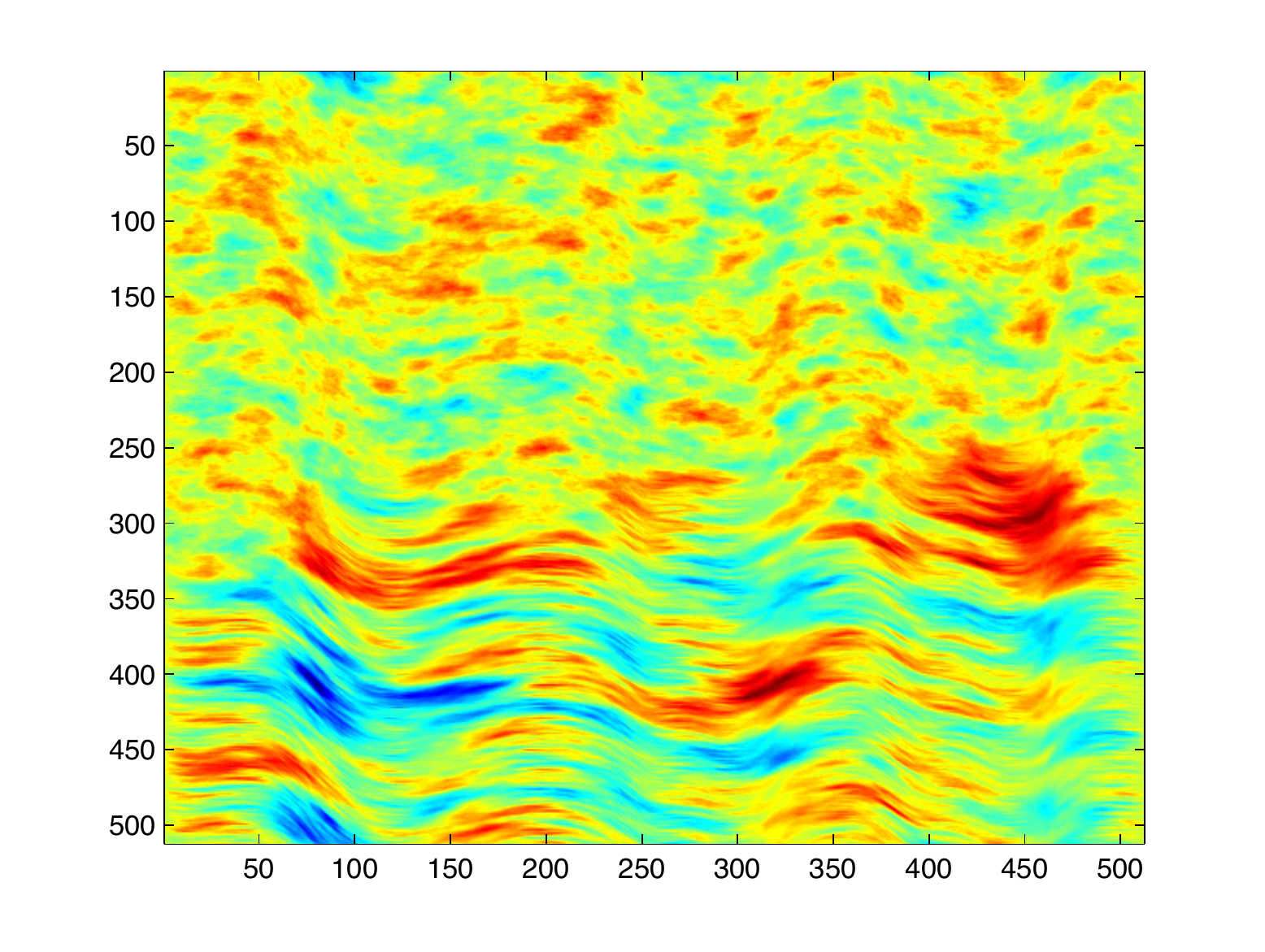}
	\end{center}
\end{figure}

There are two properties that are desirable to have in the prior model in AVA inversion. The first is being able to have different correlation length at different points in space. If a geologist have sound reasons to believe that a layer is very inhomogeneous, it may warrant putting a lower correlation length here than in a layer that is thought to be very homogeneous with very similar properties. Facilitating this is trivial - one merely lets $\kappa^2=\kappa^2(s)$ vary with space. The other property that is very desirable to have is anisotropy. Letting the correlation length vary with direction is very natural given that the layers are typically not flat but are deformed in a specific way. The SPDE resulting is the following variant of \eqref{eq_matern_stat}:
\begin{align}
	M_{\kappa,\alpha,s} x(s) =(\kappa^2(s) - \nabla \cdot \bmm{H}(s) \nabla)^{\alpha/2} x(s) = \mathcal{W}(s), \label{eq_matern_nonstat}
\end{align}
where $\bmm{A}$ is a $3\times3$ symmetric positive definite matrix defining the anisotropy angle and principal correlation length in the three directions defined by the eigenvectors of the matrix. Realisations of the stationary model and the nonstationary model is given in Figure \ref{fig_univar_stat_nonstat}. Here we have illustrated the ``layer'' flexibility mentioned above, where the top layer is isotropic, and the bottom layer is anisotropic with deformation defined by the layer.

To see how this relates to the usual approach, consider $\bmm{Q}_0=\bm{\Sigma}_0^{-1}$ and say that $m_1,m_2,m_3$ have equal Mat\'{e}rn covariance models (this includes the widely used exponential and Gaussian models), then the prior given as in \eqref{eq_prior_cov_spec} is given by the following system of stochastic differential equations:
\begin{align}
	(M_{\kappa,\alpha,s} \otimes \bmm{Q}_0)  \bmm{m}  = \bm{\mathcal{W}} \label{eq_prior_normal}
\end{align}
where $\bm{\mathcal{W}}$ is vector Gaussian spatial white noise. The experience in AVA-inversion is that at least one component of $\bmm{m}$ worse resolved than the others, with $m_1$ being resolved the best (see \cite{rab_urs_nonlinsampling} or any other article treating this problem). The obvious next question then is whether or not \eqref{eq_prior_normal} specifies the best way of lending strength to the least resolved parameters. If not, can we find better operators on the diagonal in \eqref{eq_prior_normal}, and/or replace the off-diagonals with other operators that have better properties in the inversion problem? The answer to this question is not obvious, but we investigate some alternatives and see how they perform in our inversion problem; the criterion for a better prior in the synthetic case being that $\mathbb{E}((\bmm{m_{true}}-\bmm{m_{est}^{new}})^2) < \mathbb{E}((\bmm{m_{true}}-\bmm{m_{est}^{base}})^2)$, where $\bmm{m_{est}^{base}}$ is given by the prior model \eqref{eq_prior_cov_spec}.

It is possible to replace the operator $M_{\kappa,\alpha}$ in \eqref{eq_prior_normal} by more general pseudo-differential operators. Representations of such operators in terms of its symbol are given by
\begin{align}
	(K_\sigma f)(x)=\int_{\mathbb{R}^d} \sigma(x,\xi) \hat{f}(\xi) e^{2 \pi i x \cdot \xi} d \xi,
\end{align}
where $\hat{f}$ is the Fourier transform of $f$, and $\sigma$ is the symbol of the operator. The symbol can be interpreted as defining the local spectrum of the operator. A deep theorem given in \cite{rozanov1977markov} states that a stationary random field is Markov (in the continuous sense) if and only if $\sigma^{-1}$ is a symmetric positive polynomial. Hence Markov fields are represented by differential operators. Now, if the field in question is not Markov, it is possible to approximate $\sigma$ by a rational approximation, $\sigma(x,\xi)^{-1} \approx \sigma^{-1}_\text{rat}(x,\xi) = \sum_{j=0}^k a_j(x)(2 \pi i \xi)^{j}$. To find the $a_j$s one can, for instance, use optimisation techniques. This is one way to do it, but we suspect that the time-frequency localisation of such an approach may be suboptimal, and discretization of the non-Markov operator may be better suited for time-frequency compressing approaches inducing approximate Markovity. We do not pursue these type of ideas here, but mention them as they are good candidates for future research.

\section{Systems of SPDEs -- generalising ``$\, \bmm{Q}_0$''}

It is easy to write the form the generalised approach must have. First, for $i,j=1,\ldots,3$, let
	\begin{align}
		K_{ij}=q_{ij}(\bmm{s}) (\kappa_{ij}(\bmm{s}) - \nabla \cdot \bmm{A}_{ij}(\bmm{s}) \nabla)^{\alpha_{ij}/2} \label{eq_base_mvar_nonst}
	\end{align}
	and define the following system of SPDEs
	\begin{align}
	\bmm{K}  \bmm{m}(\bmm{s})=
	\left(
		\begin{array}{ccc} 
			K_{11} & K_{12} & K_{13} \\
			K_{12} & K_{22} & K_{23} \\
			K_{13} & K_{23} & K_{33}
		\end{array} \right) \bmm{m}(\bmm{s}) = \bm{\mathcal{W}}(\bmm{s}) \label{eq_mvar_nonst}
	\end{align}
For $q_{ij}(\bmm{s})=Q^0_{ij}$ and $K_{ij}=M_{\kappa,\alpha}$ we recover the structure in the previous section with stationarity. For convenience, we call $q_{ij}(\bmm{s})$ the blending coefficients. In \cite{xiangping_mvspde}, they study the properties of this model in the stationary case, and give the link to the multivariate Mat\'{e}rn fields in \cite{gneiting2010matern}. Any choice of $K_{ij}$ defines a valid Gaussian Markov random field, both in the continuous sense and when discretized. In our treatment, we restrict ourselves to the case where $\alpha_{ij}=\alpha,\bmm{A}_{ij}(\bmm{s})=\bmm{A}(\bmm{s})$ and $\kappa_{ij}(\bmm{s})=\kappa(\bmm{s})$.

\subsection{Parametrising the blending coefficients} \label{sec_param_q0}

In general, it is both hard to interpret a local precision matrix, $\bmm{Q}_0(\bmm{s})=\{q_{ij}(\bmm{s})\}_{ij}$ defining how the individual parts of the multivariate fields is related to each other at position $\bmm{s}$, and to ensure that this matrix is positive definite.  It is much more natural to work with the inverse, namely the correlation matrix defining the local correlation of the fields, $\bm{\Sigma}_0(\bmm{s})=\bmm{Q}_0^{-1}(\bmm{s})$. The $q_{ij}(\bmm{s})$ is then simply given by the corresponding matrix elements. In the AVA inversion problem, information about correlation in different layers may come from geologists or geophysicists for who may know of phase changes when going from one layer to another in the different layers, or other, more complex phenomena. It may also come from well-logs that may contain information about such matters.

Suppose that $\bm{\Sigma}_0(\bmm{s})=\bm{\Sigma}_{0,1}$ for $\bmm{s} \in S_1 \subset \mathbb{R}^d$ and $\bm{\Sigma}_{0,2}$ for $\bmm{s} \in S_2 \subset \mathbb{R}^d$. Then we have a model that has specific correlations in one spatial region of the multivariate fields, and different correlations in another spatial region. There is obviously a transition between these two states. If the transition is discontinuous, this may be seen as a discontinuity of the correlations in the realisation of the multivariate random field, which may make sense in some situations.

\begin{figure}
	\caption{Stationary model given by \eqref{eq_prior_cov_spec}. The field looks the same wherever we are.} \label{fig_stat_mvar_geos}
	\begin{center}
		 \includegraphics[width=0.95\textwidth]{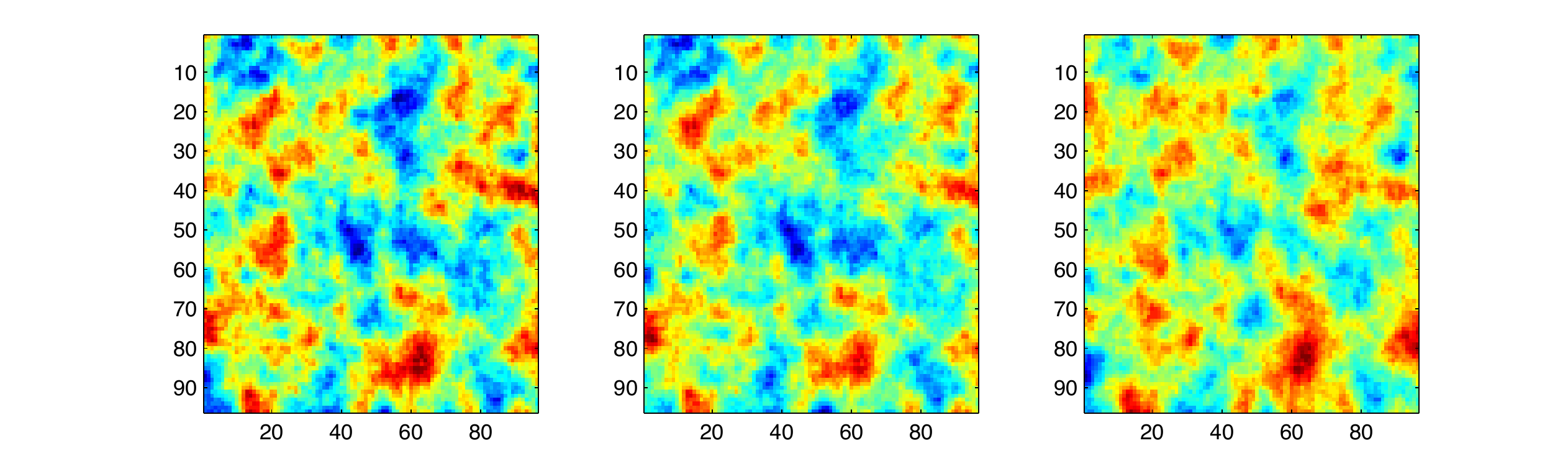}
	\end{center}
\end{figure}

\begin{figure}
	\caption{Nonstationary model with fixed $\bmm{Q}_0$. Here the bottom layer has anisotropy along the curve of the interface, and the correlation between the fields is fixed through space.} \label{fig_nonst_mvar_geos1}
	\begin{center}
	 \includegraphics[width=0.95\textwidth]{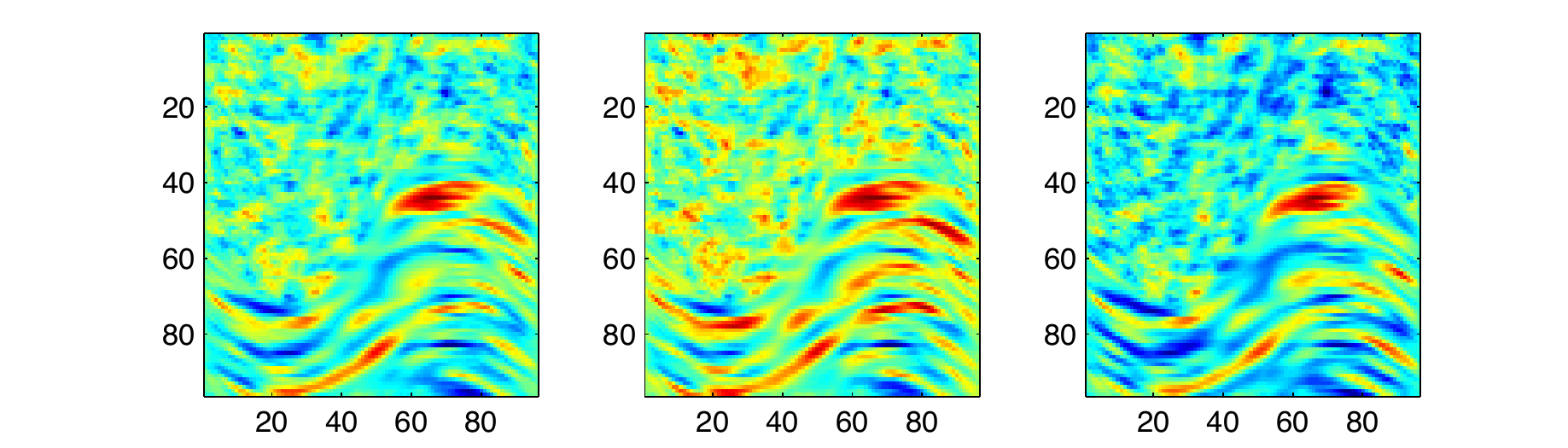}
	\end{center}
\end{figure}

\begin{figure}
	\caption{Full nonstationary model with varying $q_{ij}(\bmm{s})$ according to \eqref{eq_spd_geodes}. The bottom layer has anisotropy along the curved interface, and the correlation  between the fields changes between interfaces. In particular, the rightmost field is positively correlated to the others above the interface and negatively correlated below the interface.} \label{fig_nonst_mvar_geos2}
	\begin{center}
	 \includegraphics[width=0.95\textwidth]{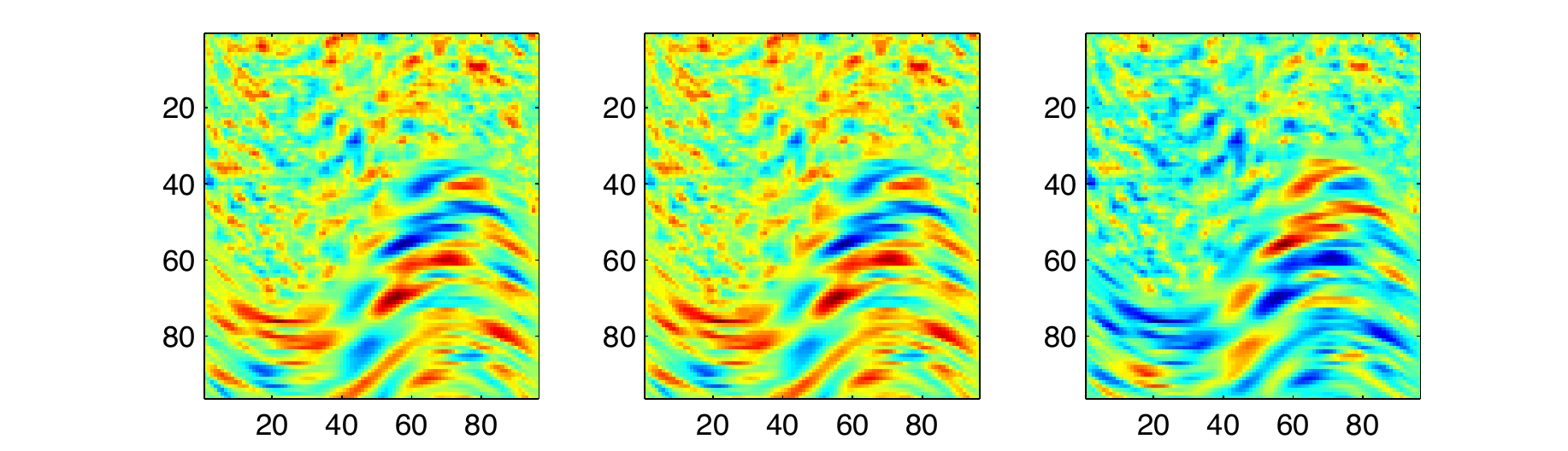}
	\end{center}
\end{figure}

In order to visualise what this means, we give realisations of the four major prior models we have discussed. In Figure \ref{fig_stat_mvar_geos}, no prior information about the geometry of the subsurface can be included. In Figure \ref{fig_nonst_mvar_geos1}, geometric information has been incorporated, but no change in the correlation between the parameters in space can be included. In Figure \ref{fig_nonst_mvar_geos2}, an example realisation from the full model is given. Pay attention to the rightmost field -- here the correlation to the other two fields changes from being positive in the top layer to being negative in the bottom layer.

\subsection{Geodesic blending}

There are obviously many ways of making a smooth transition between $\bm{\Sigma}_{0,1}$ and $\bm{\Sigma}_{0,2}$, but one key consideration is that $\bm{\Sigma}_0(\bmm{s})$ must remain positive definite for all $\bmm{s}$ in some transition domain $S_T$. One thing is certain - it is not necessarily enough to let the off-diagonals element in $\bm{\Sigma}_{0,1}$ change linearly in $\mathbb{R}^3$ to the corresponding off-diagonal elements in $\bm{\Sigma}_{0,2}$.

A very natural way of making such a transition between $\bm{\Sigma}_{0,1}$ and $\bm{\Sigma}_{0,2}$ is by considering geodesics on the manifold of symmetric positive definite matrices, denoted $\mathbb{P}_d$. The natural metric on this space has a reasonable statistical interpretation, closely related to information entropy and Kullback-Leibler divergence, and an accessible account for the theory is given in \cite{bhatia2007positive}. Different treatments are given in \citep{ohara1996dualistic,hiai2009riemannian}. For completeness, we give a small account of the definition and properties we need related to this manifold. This exposition is based on \cite{hiai2009riemannian,bhatia2007positive}.

The Boltzmann entropy of the Gaussian distribution \eqref{eq_gauss_simple}, defining an information potential, is given by
\begin{align}
	B(p(\bmm{x}|\bmm{Q},\bm{\mu}_x))=B(\bmm{Q})=\frac{1}{2}\log \det \bm{\Sigma} + C,
\end{align}
where $C$ is an arbitrary constant and $\bm{\Sigma}=\bmm{Q}^{-1}$ is any positive definite matrix. The Riemannian metric based on this information potential is the Hessian 
\begin{align}
	\text{g}_{\bmm{Q}}(\bm{\mathrm{H}},\bm{\mathrm{M}})=\frac{\partial^2}{\partial s \partial t} \Bigg|_{s=0,t=0} B(\bmm{Q}+s\bm{\mathrm{H}}+t\bm{\mathrm{M}})=\text{tr} \, \bmm{Q} \bmm{H} \bmm{Q}\bmm{K} ,
\end{align}and
where $\bmm{H},\bmm{S} \in \mathbb{S}_d$, the tangent space of symmetric matrices, $\mathbb{S}_d = \{\bm{\mathrm{V}} \in \mathbb{R}^{d \times d} |  \bm{\mathrm{V}} = \bm{\mathrm{V}}^T\}$. This defines the line element
\begin{align}
	ds = \left(\text{tr}  \left[ (\bmm{Q}^{-1/2} d\bmm{Q} \bmm{Q}^{-1/2})^2 \right]\right)^{1/2}.
\end{align}
Hence, if we have a curve in $\mathbb{P}_d$, i.e. $\gamma: [a,b] \to \mathbb{P}_d$, its length can be calculated as
\begin{align}
	L(\gamma)=\int_a^b \left( \text{tr} \left[ (\gamma(t)^{1/2} \gamma'(t) \gamma(t)^{1/2})^2 \right] \right)^{1/2} dt
\end{align}
A nice property that follows from this is that lengths of curves are invariant under congruence transformations. That is, if $g(t)=\bmm{X}^T \gamma(t) \bmm{X}$, $L(\gamma)=L(g)$. The geodesic, the curve with minimal length, between two matrices, $\bmm{A}$ and $\bmm{A}$ can from this be deduced to be
\begin{align}
	\text{g}_{\bmm{A},\bmm{B}}(t) = \bmm{A} \#_t \bmm{B}=\bmm{A}^{1/2}\left(\bmm{A}^{-1/2} \bmm{B} \bmm{A}^{-1/2} \right)^t \bmm{A}^{1/2}, \quad t \in [0,1]. \label{eq_spd_geodes}
\end{align}
Obviously, $\text{g}_{\bmm{A},\bmm{B}}(0)=\bmm{A}$ and $\text{g}_{\bmm{A},\bmm{B}}(1)=\bmm{B}$. It is this curve we use when we go from $\bmm{A}=\bmm{Q}_{0,1}=\bmm{\Sigma}_{0,1}^{-1}$ to $\bmm{B}=\bmm{Q}_{0,2}=\bmm{\Sigma}_{0,2}$ in different discrete structures in our prior model, and this ensures that we are within the realm of positive definite matrices in a natural way. Noting that $(\bmm{A} \#_t \bmm{B})^{-1} = \bmm{A}^{-1} \#_t \bmm{B}^{-1}$, we see that it is unproblematic to work with precision matrices rather than covariance matrices. Integrating $\text{g}_{\bmm{A},\bmm{B}}(t)$ yields the distance between the two matrices,
\begin{align}
	d_{\mathbb{P}_d}(\bmm{A},\bmm{B})=\int_0^1 \text{g}_{\bmm{A},\bmm{B}}(t)= \left( \text{tr} \left[ (\log \bmm{A}^{-1/2} \bmm{B} \bmm{A}^{-1/2})^2 \right] \right)^{1/2}.
\end{align}

A potential drawback of using this strategy is that if $\bmm{Q}_{0,1},\bmm{Q}_{0,2}$ are correlation matrices, and what you want is a continuum of correlation matrices, $\text{g}_{\bmm{Q}_{0,1},\bmm{Q}_{0,2}}(t)$ are not correlation matrices for $t \in(0,1)$. It is possible to correct for this by using geodesics on the submanifold of correlation matrices in $\mathbb{P}_d$. In practice, however, $\text{g}_{\bmm{Q}_{0,1},\bmm{Q}_{0,2}}(t)$ are very close to being correlation matrices in most cases. We do not have any counterexamples.

\subsubsection{Fuzzy interfaces} 

In some situations, we may actually have a hard interface in our multivariate field, but even in this situation, experts may place the interface incorrectly, which may lead to imprecise interpretation of the field. The geodesic blending strategy discussed in the previous section gives us a way to handle this situation in a specific way: the blending range may serve as quantifying the uncertainty or fuzziness of this interface. This range may then be estimated based on realisations of the field, possibly requiring a strong prior for identifiability.

\begin{figure}
	\caption{Illustration of blend range. Guessed interface (left), true interface (right), blend range illustrated in grey (bottom)} \label{fig_blend_range_geodesic}
	\begin{center}
		\includegraphics[width=0.45\textwidth]{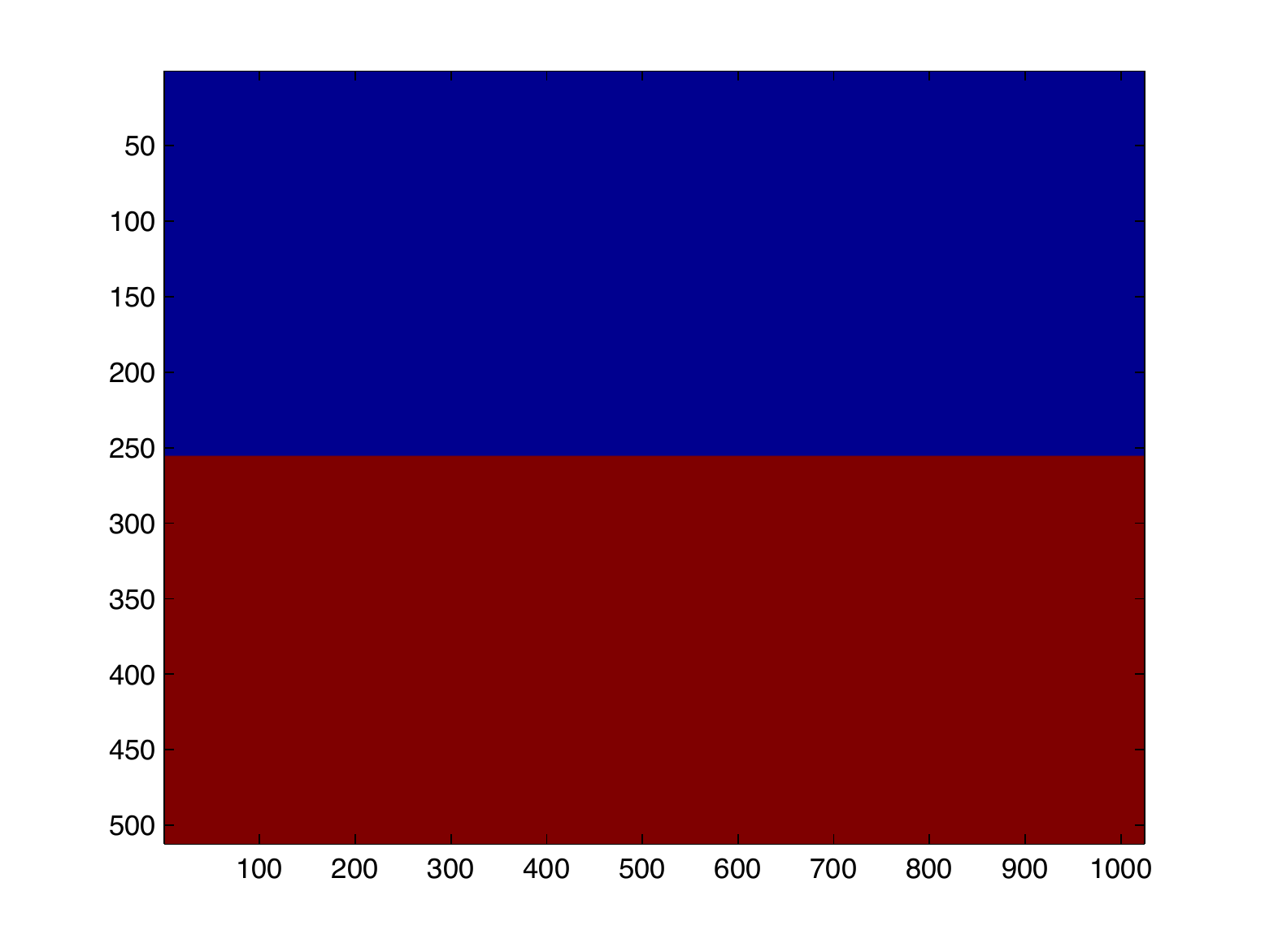} \includegraphics[width=0.45\textwidth]{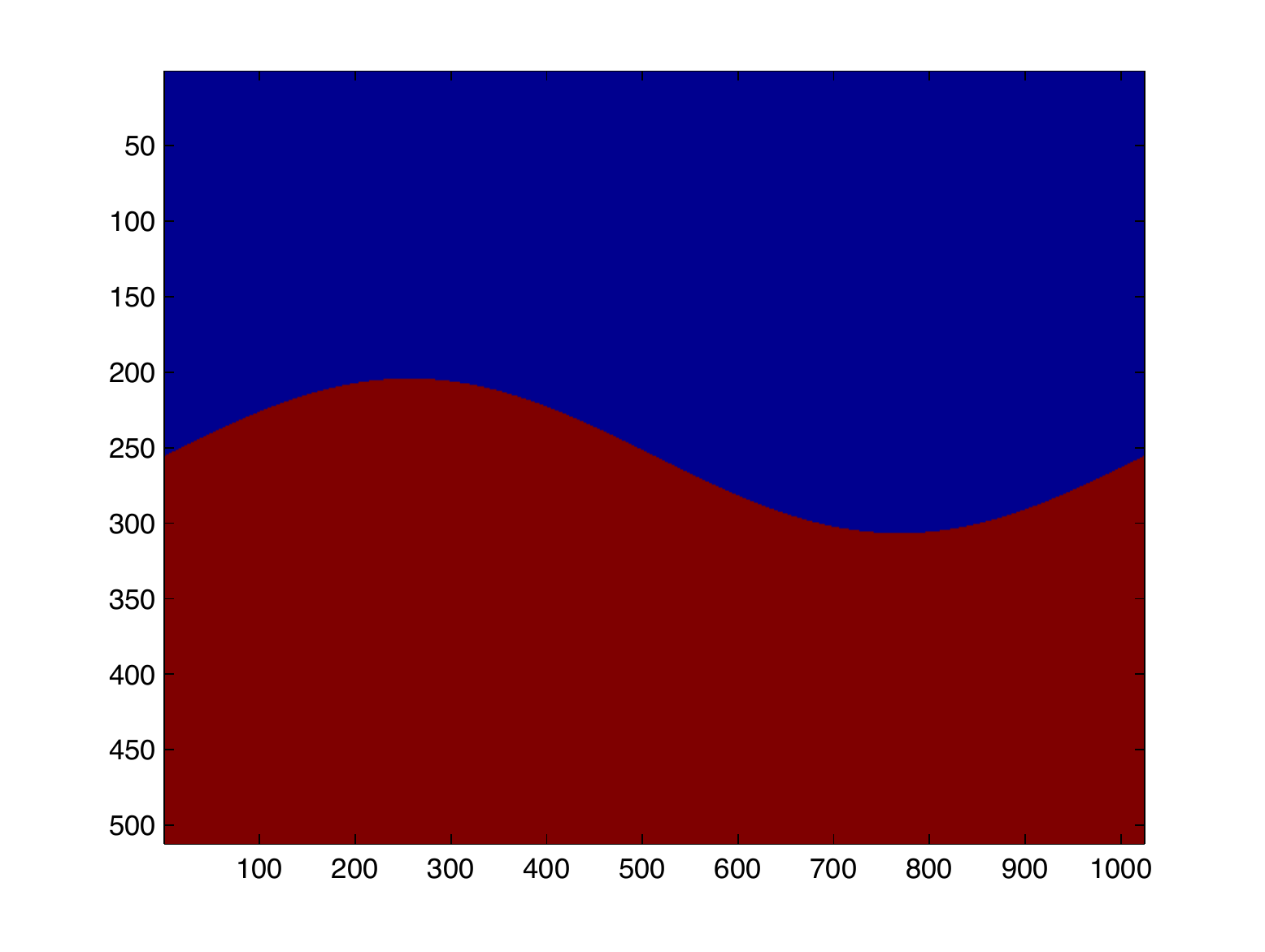} \\
		\includegraphics[width=0.45\textwidth]{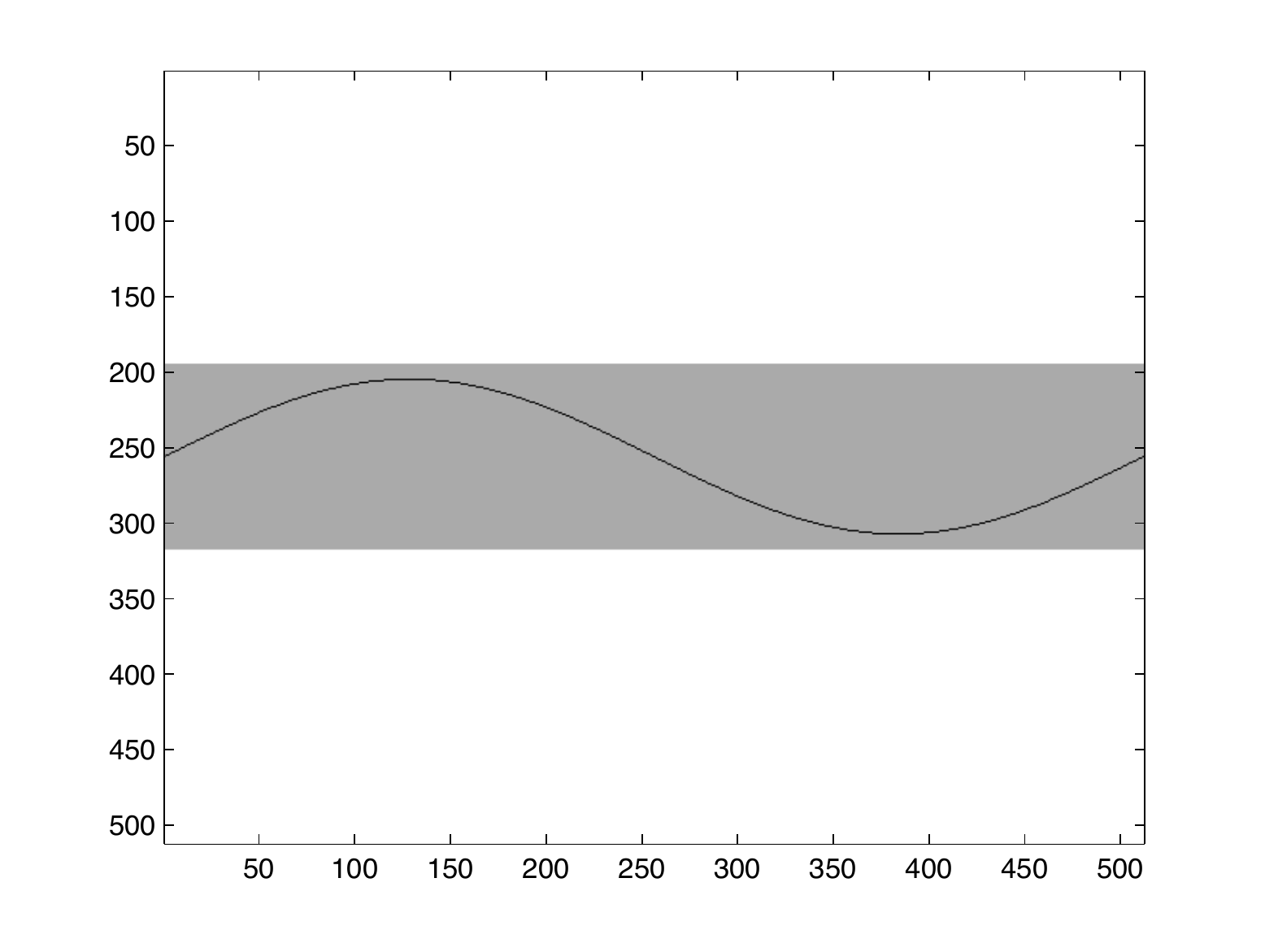}
	\end{center}
\end{figure}

To illustrate this, suppose that an expert says that the interface is as in the upper left part of Figure \ref{fig_blend_range_geodesic}, while the real interface is given on the right. The bottom illustration in Figure \ref{fig_blend_range_geodesic} shows what the geodesic range should be in this case (grey area) -- it should cover the true interface properly, showing that there actually is a fair amount of uncertainty in the placement of the interface. In Section \ref{sec_parest_spde_ava}, we investigate whether this range may be estimated purely from data or if a strong prior on the range is needed. It is, of course, possible to combine this idea with procedures for actually estimating the interface, but, as always, this increases the complexity of the model that is to be estimated. Additionally, the blend range may easily confound with potential parameters needed to estimate the actual location of the interface.

\subsection{Modified parametrisations}

There are many ways to modify the parametrisation described above to reduce parametrisation demand or incorporate different flexibility. A possible way to reduce the parametrisation demand further is to do the modeling in the Cholesky domain. This is a simplification, but it is one we believe should increase interpretability and possibly estimation properties. To motivate this approach, consider the following: Suppose that the Cholesky factorisation of $\bmm{Q}_0$ is given by $\bmm{Q}_0=\bmm{L}_0 \bmm{L}_0^T$, and that $\bmm{Q}_\text{space} = \bmm{B}^s_{1} (\bmm{Q}^s_{2})^*$, for some matrices $\bmm{B}^s_1,\bmm{B}^s_2$. Generating the matrices $\{\bmm{B}^s_{i}\}_{1,2}$ can for instance be done by using $\alpha_s = \alpha/2$ in \eqref{eq_matern_stat} and discretizing this operator, but there exist many other factorisations that may behave in better way for the problem at hand. By a Kronecker product identity, $\bmm{Q}_\text{space} \otimes \bmm{Q}_0 = (\bmm{B}^s_1 \otimes \bmm{L}_0) ((\bmm{B}^s_2)^* \otimes \bmm{L}_0^T)$. The intuition stemming from this identity carries over to the more general case in a natural way: Let $l_{ij}(\bmm{s})$ be entry $i,j$ of the Cholesky factor of the matrix $\{q_{ij}(\bmm{s})\}_{ij}$ locally, and define locally operators that will correspond to some square root of its original form in \eqref{eq_base_mvar_nonst}. It is possible to define the operators in such a way that we get back \eqref{eq_mvar_nonst}, but this is of minor concern in practice as long as we get the interpretability we want. This is remniscient to the triangular approach mentioned in \cite{xiangping_mvspde}.

\section{Parameter estimation and conditional expectation} \label{sec_parest_spde_ava}

In order to show that our proposed model is useful with confidence in the realm of seismic AVA inversion, we must show that estimation of hyper-parameters in the prior model is feasible and that the conditional expectation, $\mathbb{E}(\bmm{m}|\bmm{d},\bm{\theta})$, is better than in the simpler model. A natural way to see if the hyper-parameters are identifiable is to simulate from the prior fields and do maximum likelihood estimates on these. If this works well, one may go one level higher and assume noisy observations of the form
\begin{align}
	\bmm{y} = \bmm{WA}\bmm{x} + \bm{\epsilon},
\end{align}
where $\bmm{W}$ denotes a convolution matrix defined by the wavelet, and $\bmm{A}$ denotes the reflectivity matrix, and $\bm{\epsilon} \sim \mathcal{N}(\bm{0},\bmm{I})$. It is also here possible to do maximum likelihood estimates. For more information on estimating this type of model, consult your favourite treatise that discusses latent (Gaussian) models for, e.g. \cite{rasmussen_gauss_machine,rue_gmrf,cressie2011statistics}. In treating this estimation problem, we use the simpler anisotropic model where the correlation changes from positive to negative at an interface defined by a straight line. It is, of course, possible to estimate the geometry as well, but this is beyond the scope of this text.

When estimating the $q_{ij}$, supposing it changes between layers, we must impose constraints to enforce the interpretability we want -- namely that of its local inverse being the correlation matrix of the multivariate field at that point. Now, the matrix consisting of ones on its diagonal with three free parameters off its diagonal uniquely specifies these constraints through its eigenvalues: they must all be greater than zero. Hence we have three constraints, depending only on the off-diagonal elements of the local correlation matrix. The same type of restriction would apply if we were to use general local $3\times 3$ covariances instead. In that scenario, however, the three constraints would depend on six parameters instead of three.  In this section, we will denote the different models as follows
\begin{enumerate}
	\item Model 1 is the simple stationary $\bmm{Q}_0 \otimes \bmm{Q}_\text{space}$ as in \eqref{eq_prior_cov_spec}
	\item Model 2 is stationary in space using the extended $q_{ij}(\bmm{s})$ parametrisation as in Section \ref{sec_param_q0}, equation \eqref{eq_mvar_nonst}, using interpretability constraints
	\item Model 3 is nonstationary in space and using the extended $q_{ij}(\bmm{s})$ parametrisation as in Section \ref{sec_param_q0}, equation \eqref{eq_mvar_nonst}. Additionally, we use a blending of two correlation matrices at the interface, so that the correlation change is not discontinuous.
\end{enumerate}

\subsection{Identifiability}

\begin{figure}
	\caption{Maximum likelihood density estimates for correlation parameters, $\kappa^2$ and $\tau^2$ using direct observations of 200 fields} \label{fig_direct_ksdens}
	\begin{center}
	\includegraphics[width=0.45\textwidth]{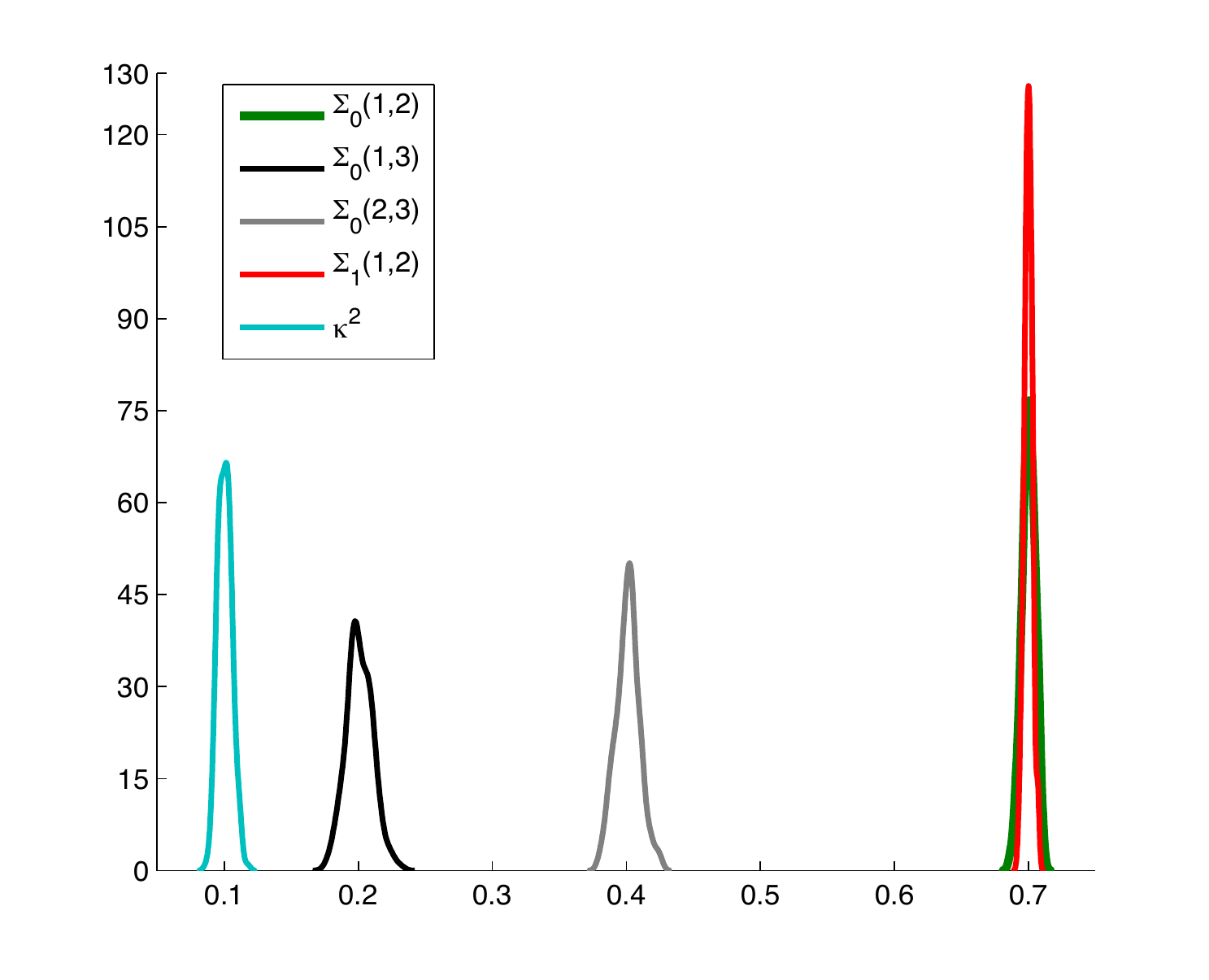} \includegraphics[width=0.45\textwidth]{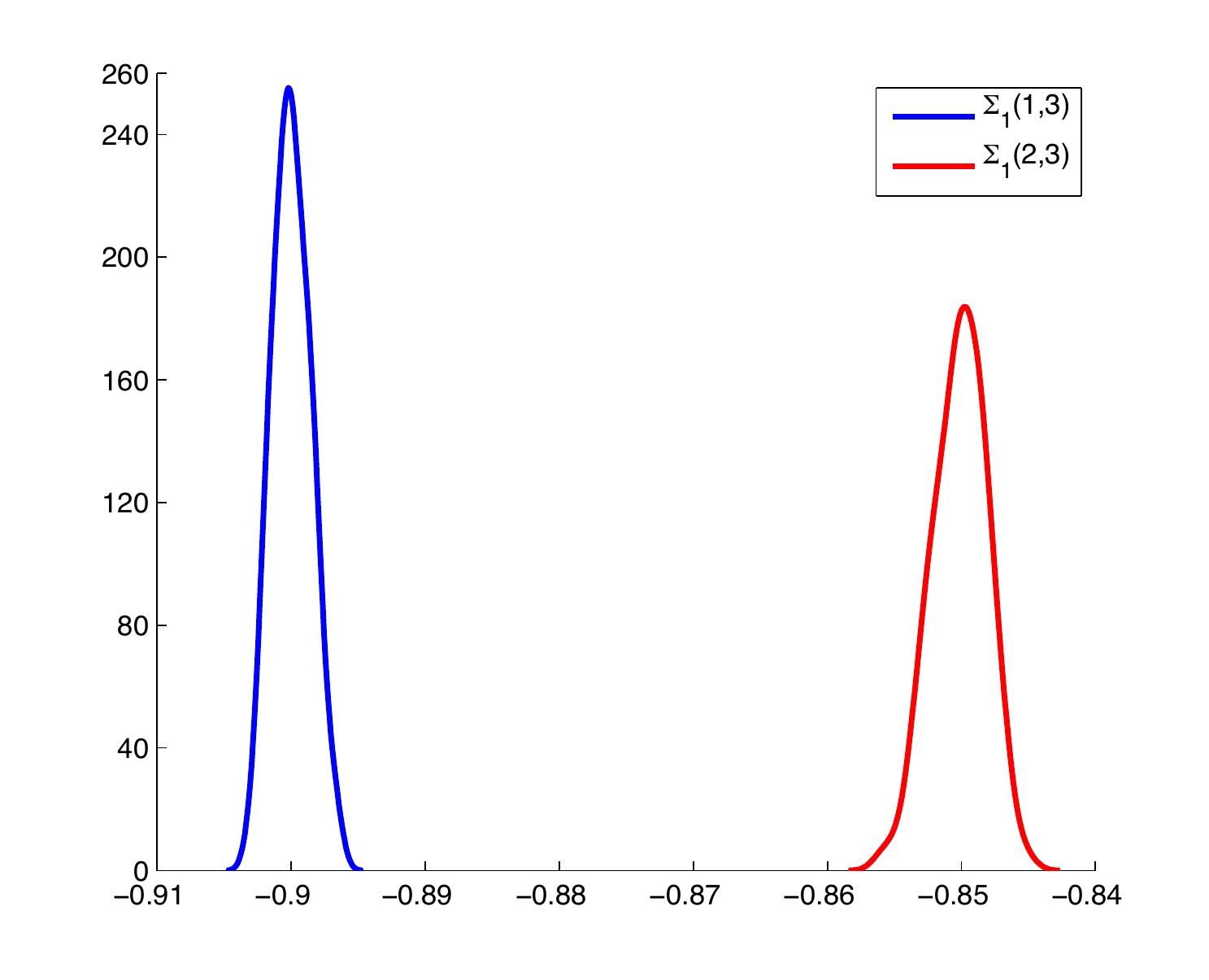}\\
	\includegraphics[width=0.45\textwidth]{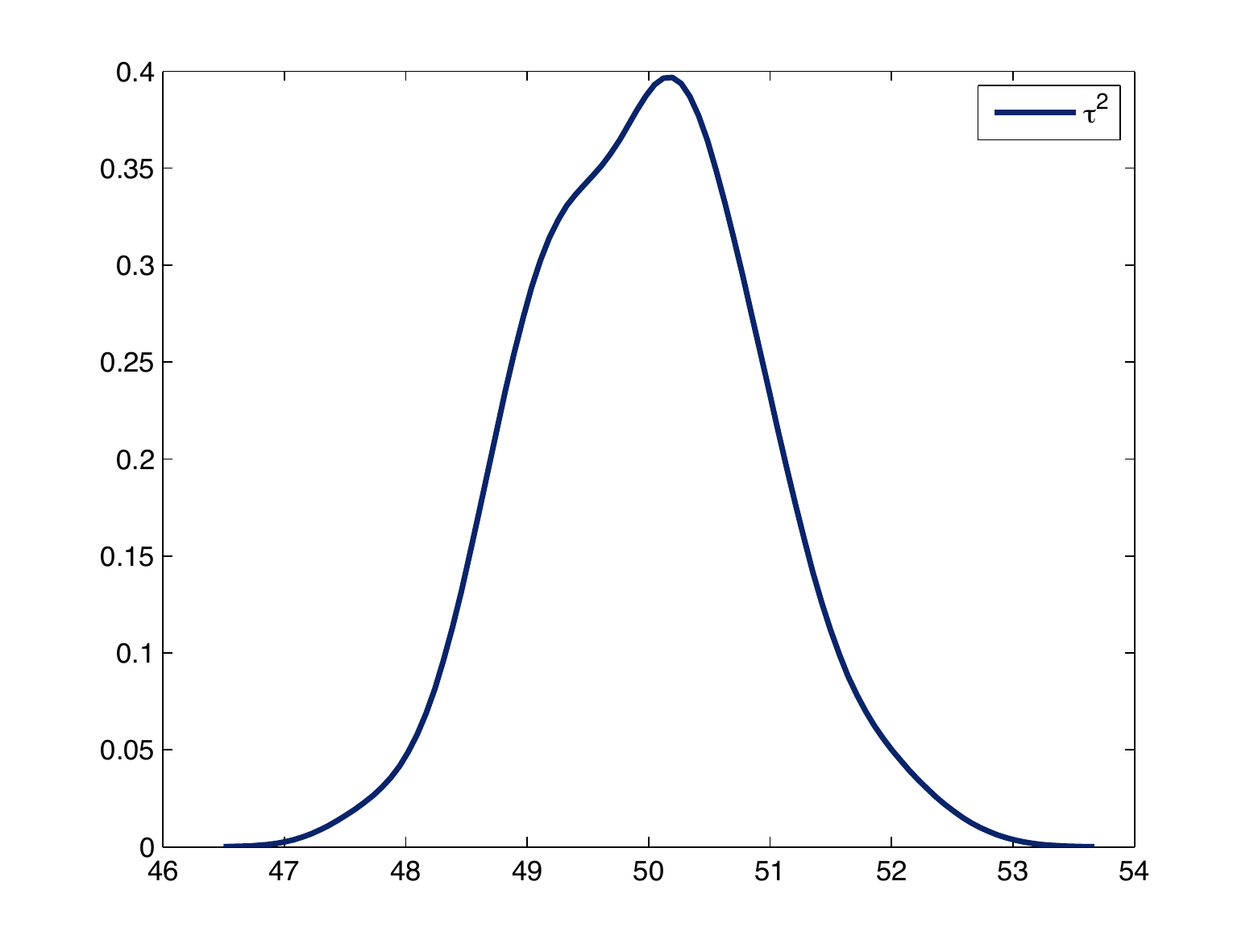}
	\end{center}
\end{figure}

We show that the parameters in $\bmm{\Sigma}_0,\bmm{\Sigma}_1$ are identifiable through simulation. To do this, we simulate from many multivariate fields and estimate the parameters by maximum likelihood. If the estimated maximum likelihood density -- which is estimated from several realisations -- is unimodal, the parameters are identifiable. Suppose that $\bm{\Sigma}_0,\bm{\Sigma}_1$ are given by
\begin{align}
	\bmm{\Sigma}_0 = \left( \begin{tabular}{ccc} 1 & 0.7 & 0.2  \\ & 1 & 0.4 \\ & & 1 \end{tabular} \right), \quad 
	\bmm{\Sigma}_1 = \left( \begin{tabular}{ccc} 1 & 0.7 & -0.9  \\ & 1 & -0.85 \\ & & 1 \end{tabular} \right),
\end{align}
using $\kappa^2=0.1$, and $\tau^2 \bmm{Q}$, with $\tau^2=50$. Using 200 realisations from the field, we get maximum likelihood density estimates for the parameters -- these are illustrated in Figure \ref{fig_direct_ksdens}. For obtaining the parameters, we used a quasi-Newton method with initial correlation parameters being zero. Judging from this figure, since all density estimates are unimodal, all parameters seem to be identifiable.

\begin{figure}
	\caption{Maximum likelihood density estimates for correlation parameters, $\kappa^2$ and $\tau^2$ using indirect observations of 600 fields} \label{fig_inDirect_ksdens}
	\begin{center}
	\includegraphics[width=0.45\textwidth]{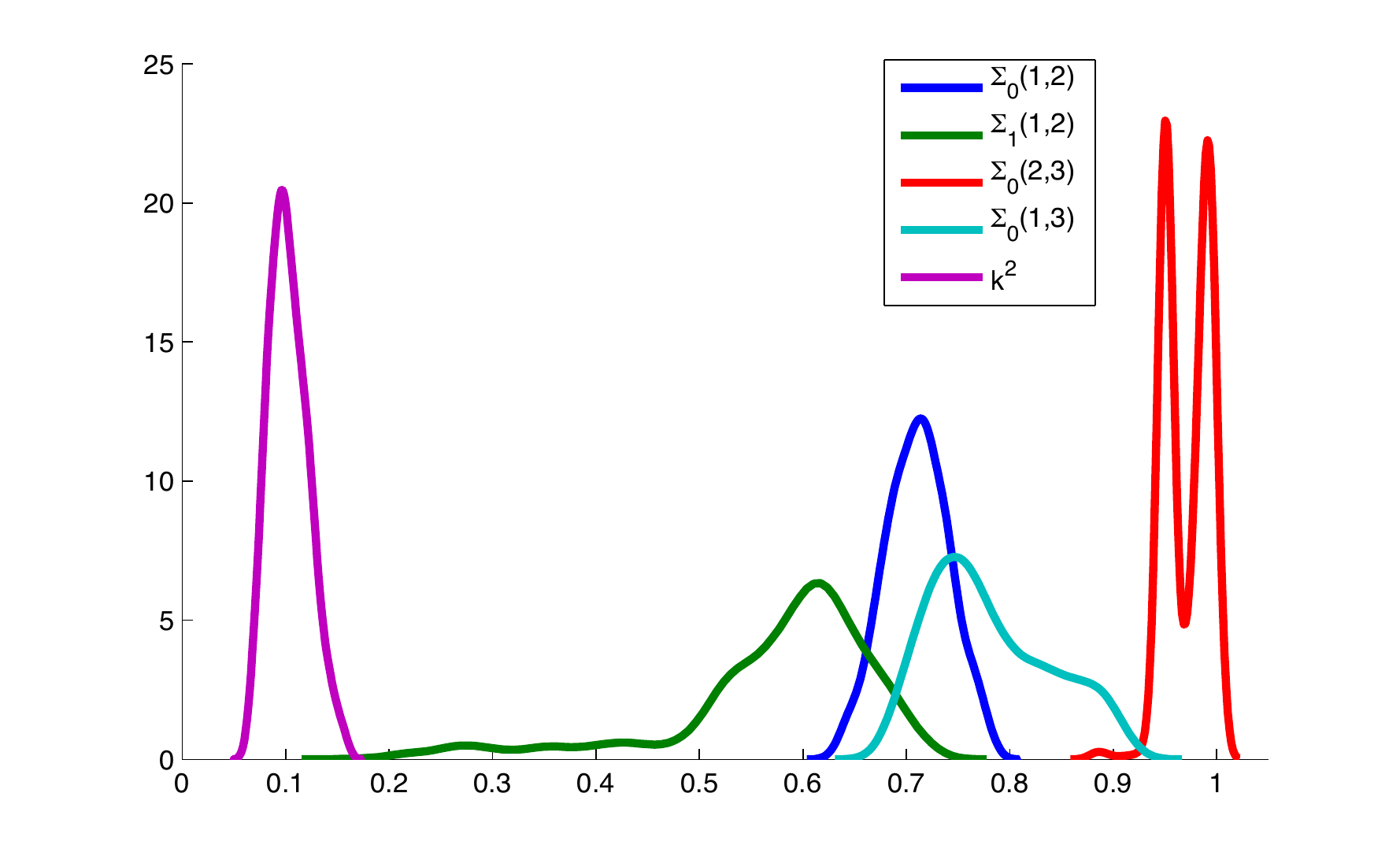} \includegraphics[width=0.45\textwidth]{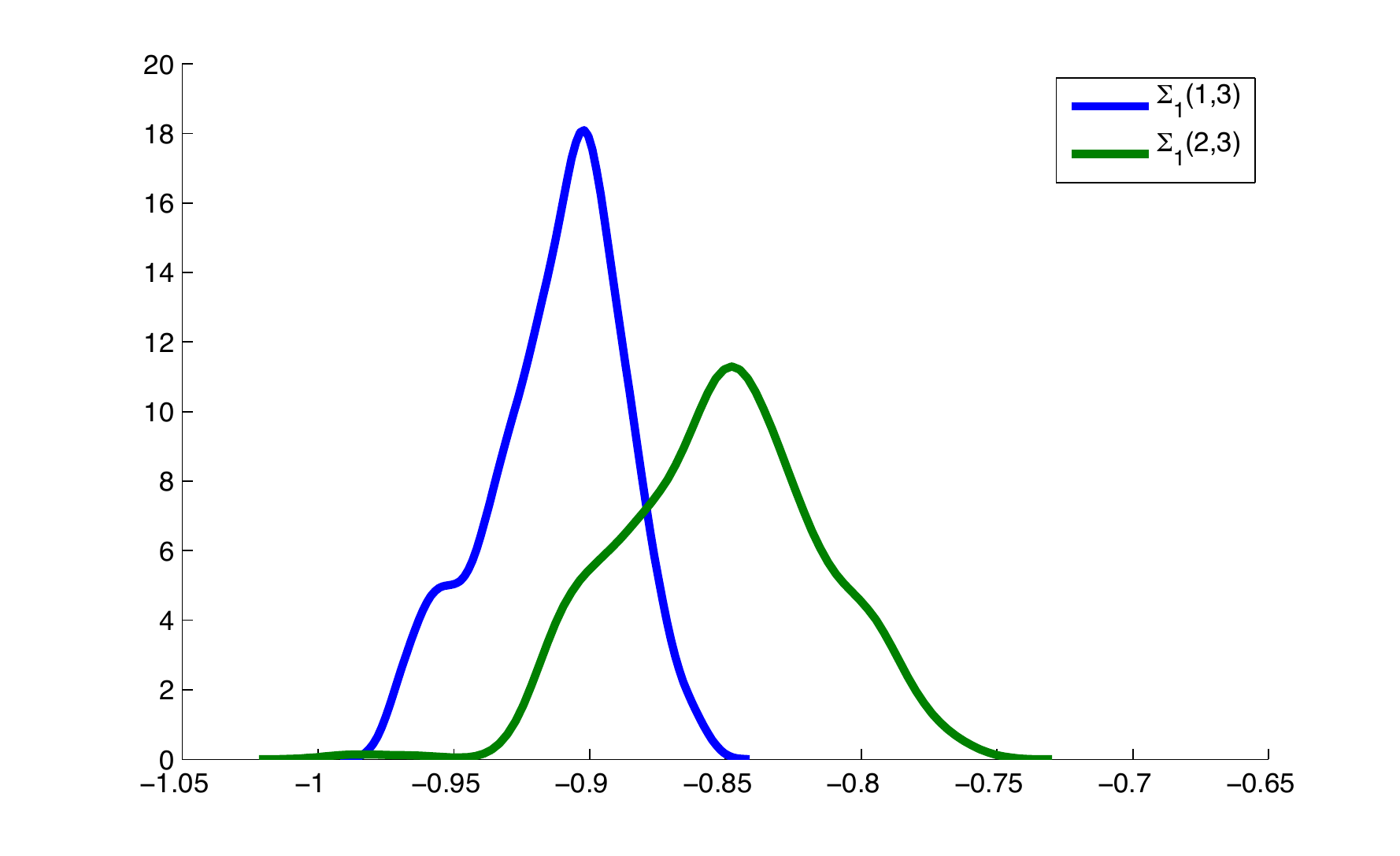}\\
	\includegraphics[width=0.6\textwidth]{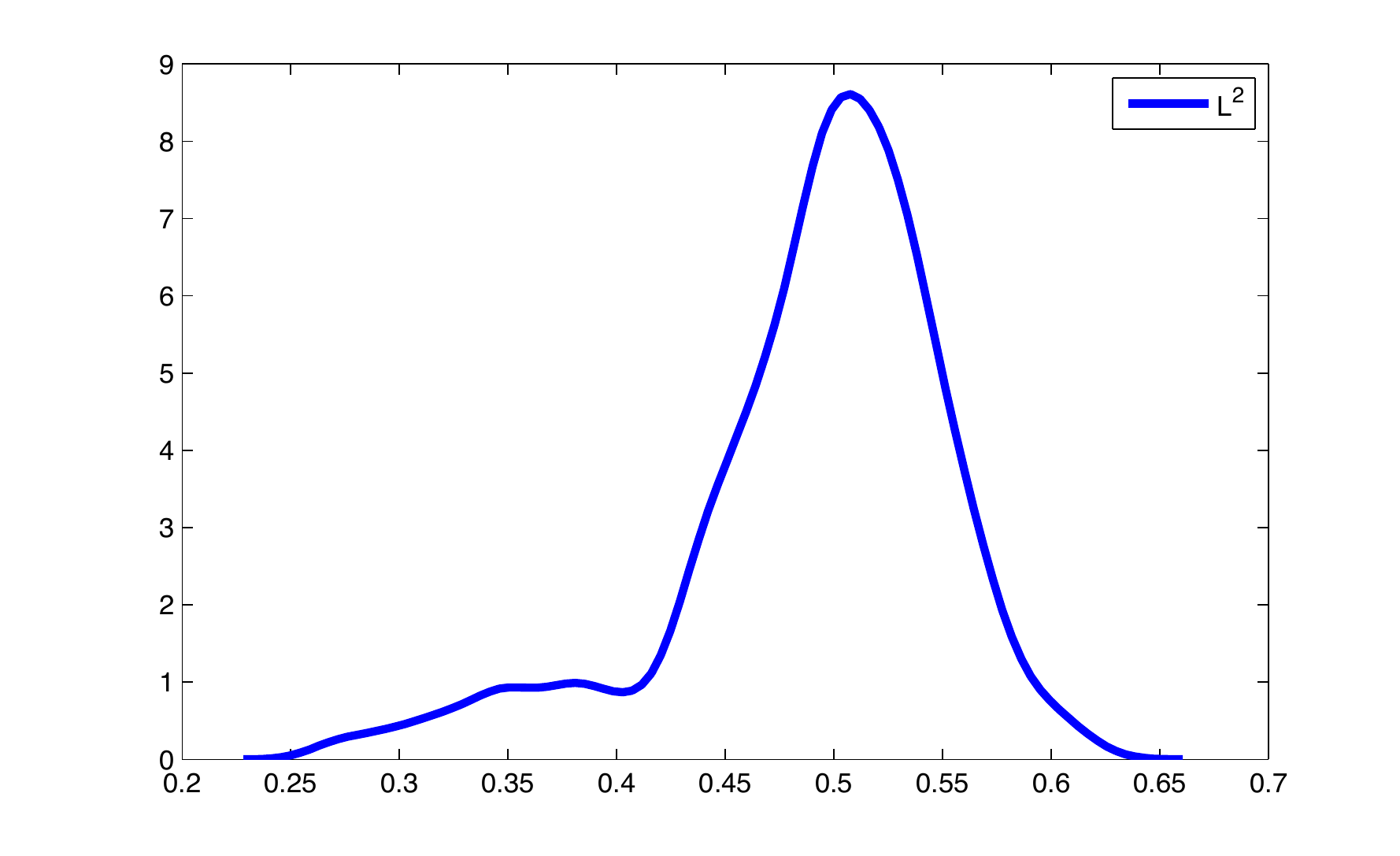}
	\end{center}
\end{figure}

In the case where we have noisy observations, we use profile likelihood to estimate the noise level, $\sigma^2$, and a quasi-Newton method to estimate $\lambda^2=\tau^2 \sigma^2$, $\kappa^2$ and the correlation parameters. In this case we used the correlation matrices
\begin{align}
	\bmm{\Sigma}_0 = \left( \begin{tabular}{ccc} 1 & 0.7 & 0.6  \\ & 1 & 0.95 \\ & & 1 \end{tabular} \right), \quad 
	\bmm{\Sigma}_1 = \left( \begin{tabular}{ccc} 1 & 0.75 & -0.9  \\ & 1 & -0.85 \\ & & 1 \end{tabular} \right).
\end{align}
In Figure \ref{fig_inDirect_ksdens} the corresponding estimates for a hidden field is given. Of course, it is much more difficult in this situation, which is reflected through the broad distributional tails in the figure. Overall, however, the estimates seem to recover the true values quite well. One odd observation is the bimodality of $\Sigma_0(2,3)$. We believe it may come from observing rather small fields, from a $64 \times 64$-grid, and that it may go away for larger ones. The values over one on the left part of the figure are artefacts coming from using a kernel smoother for estimating the density.

\subsection{Conditional expectation}

The real test on whether it is wise or not to use this advanced parametrisation of the model is essentially the reconstruction problem: based on noisy observations, are we able to reconstruct the original fields with higher fidelity? 

In the following subsections, we give several reconstructions examples, and we use two observations schemes. The first one is based on identity observations with i.i.d. noise, and the second is based on the AVA model, giving the observation matrix $\bmm{WA}$ followed by i.i.d. noise.

\subsubsection{Identity observations}

First, we present a reconstruction example where the observation matrix is the identity, followed by iid. noise, and with two signal-to-noise ratios. One with $\lambda^2=50$ and one with $\lambda^2=0.5$.  For these two models, we use the following true correlation matrices
\begin{align}
	\bmm{\Sigma}_0 = \left( \begin{tabular}{ccc} 1 & 0.99 & 0.99  \\ & 1 & 0.99 \\ & & 1 \end{tabular} \right), \quad 
	\bmm{\Sigma}_1 = \left( \begin{tabular}{ccc} 1 & -0.99 & -0.99  \\ & 1 & 0.99 \\ & & 1 \end{tabular} \right).
\end{align}

\begin{figure}
	\caption{Kriging for identity observations with signal-to-noise ratio $1/50$. True parameters (left), kriging using true Model 2 (center), using Model 1 (right)} \label{fig_krig_id_obs_sign50}
	\begin{center}
		\includegraphics[width=0.95\textwidth]{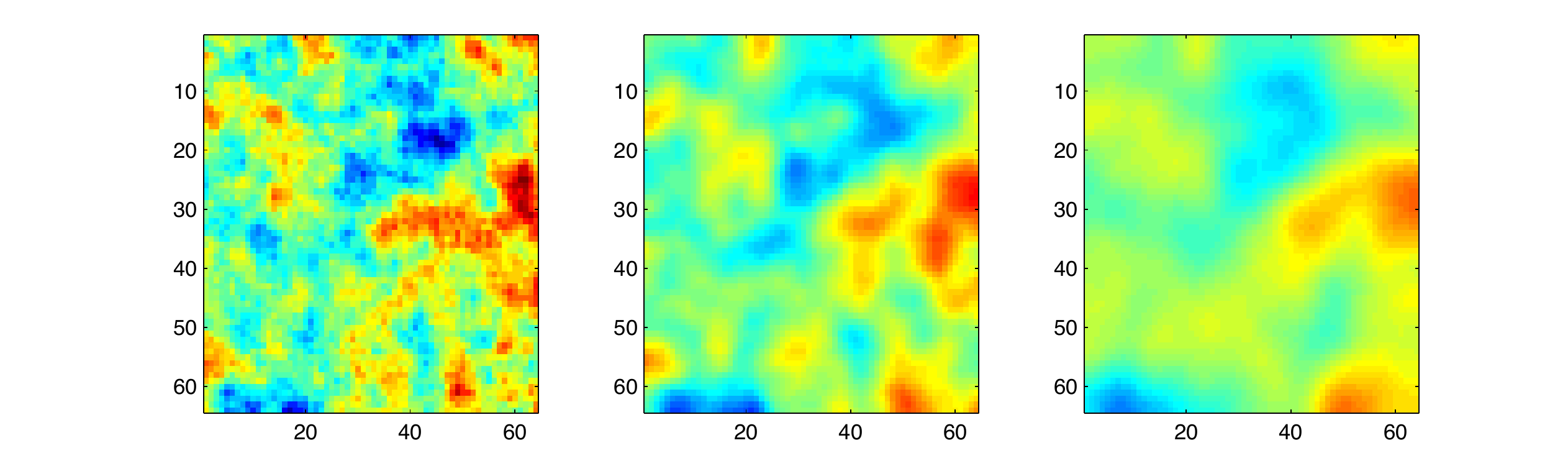}
	\end{center}
\end{figure}

In Figure \ref{fig_krig_id_obs_sign50}, we illustrate reconstruction of the first of the three fields with signal-to-noise ratio $1/50$, using a flat interface and identity observations. I.e. the field true field is generated by Model 2, followed by i.i.d. noise. A priori we believe one of the worst situations for estimating Model 1, as correlations change very much from structure to structure and the noise level is very high. The likelihood function in the situation with high noise levels appears very flat, requiring high accuracy and many iterations in the optimisation scheme to give consistent estimates. For $\lambda^2=50$, $\|\mathbb{E}^\text{M2}(\bmm{x}|\bmm{y},\bm{\theta})- \bmm{x} \|_2/ \|\bmm{x}\|_2 =0.526$ for Model 2 and $\|\mathbb{E}^\text{M1}(\bmm{x}|\bmm{y},\bm{\theta})- \bmm{x} \|_2/ \|\bmm{x}\|_2 =0.674$ for Model 1. The first field is chosen, as for the correlation matrices defined for this, the first field is the one with changing correlation between interfaces, relative to the others. The main effect we see in this comparison is that the level of the reconstructed field using the Model 1 does not completely reach up to the true levels -- we believe this can be attributed to a flattening effect arising from the sum of the two correlations in the different layers being zero.

\begin{figure}
	\caption{Kriging for identity observations with signal-to-noise ratio $1/0.5$, field 1. True parameters (left), kriging using Model 2 (center), using Model 1 (right)} \label{fig_krig_id_obs_sign05}
	\begin{center}
		\includegraphics[width=0.95\textwidth]{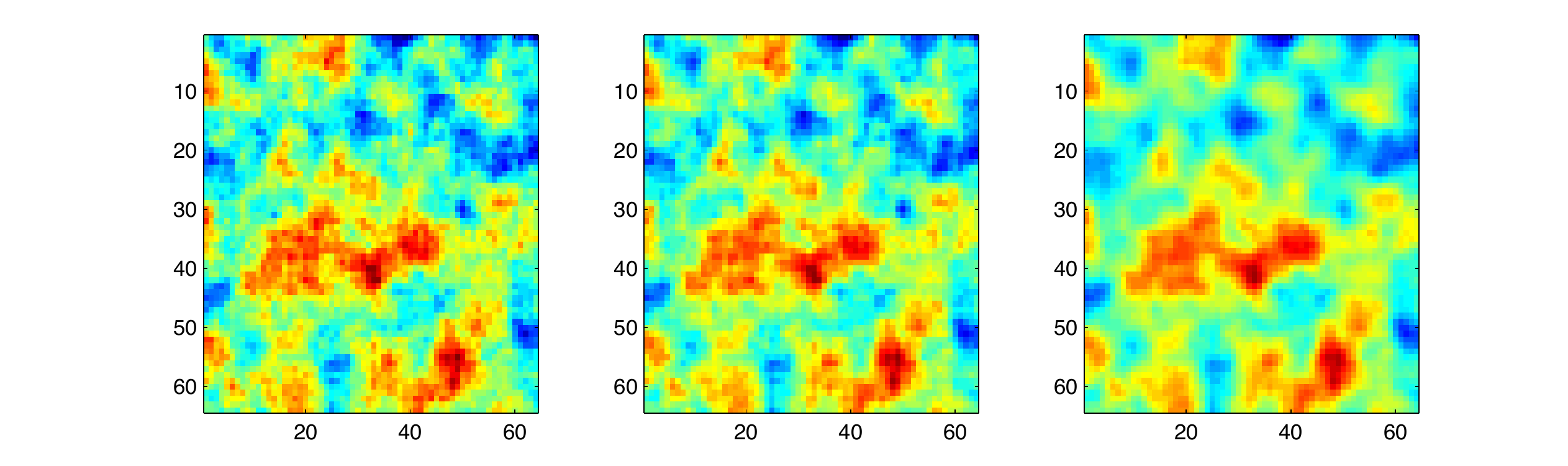}
	\end{center}
\end{figure}

\begin{figure}
	\caption{Kriging for identity observations with signal-to-noise ration $1/0.5$, field 2. True parameters (left), using true model (center), using model defined by \eqref{eq_prior_normal} (right)} \label{fig_krig_id_obs_sign05_2}
	\begin{center}
		\includegraphics[width=0.95\textwidth]{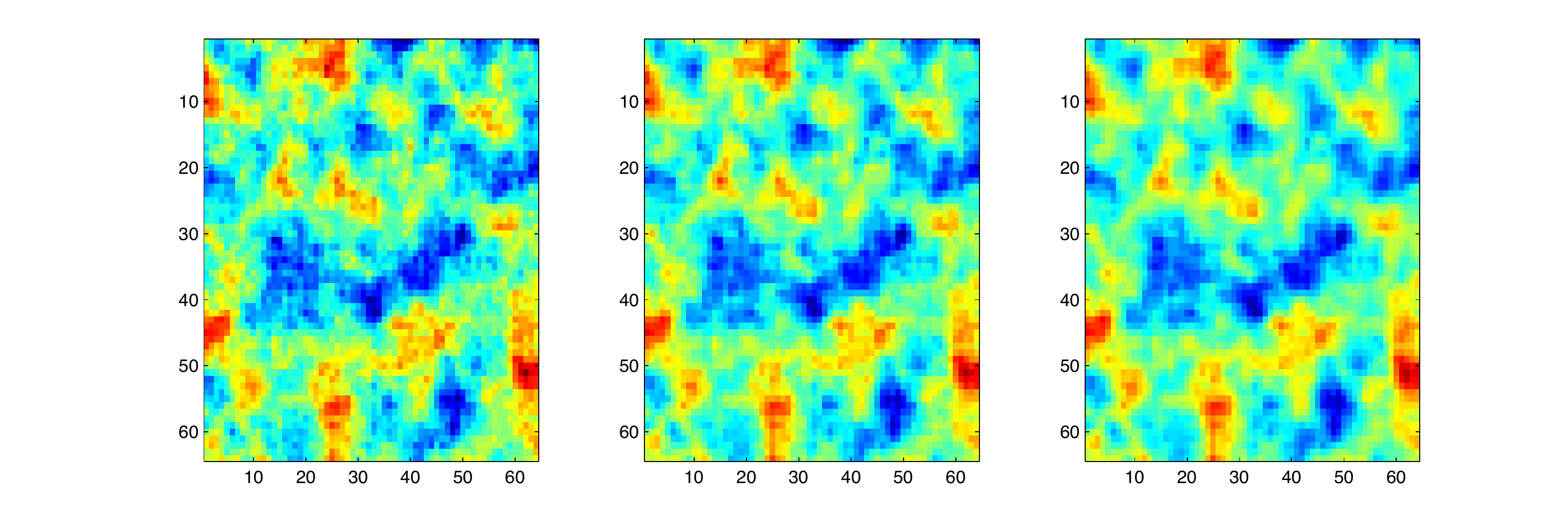}
	\end{center}
\end{figure}

In the second comparison we generate fields and observations from Model 2, with $\lambda^2=0.5$. Here a different effect is more prominent -- we see that the reconstructed field on the right in Figure \ref{fig_krig_id_obs_sign05}, i.e. the reconstruction based on using Model 1, is smoother and does not exhibit as much of the jaggedy effect of the true surface compared to the field in the middle. A consistent property when estimating the Model 1 is that $\kappa^2$ seems to be underestimates, leading to a larger range and hence smoother reconstruction. One may think that this smoothing effect of the field on the right in Figure \ref{fig_krig_id_obs_sign05}, but for comparison, we also include reconstructions of the second field, depicted in Figure \ref{fig_krig_id_obs_sign05_2}. Here the mentioned smoothing effect is not as present as in Figure \ref{fig_krig_id_obs_sign05}. Hence, we believe that this is an effect induced by the changing correlations. In Figure \ref{fig_krig_id_obs_sign05}, $\|\mathbb{E}^\text{M2}(\bmm{x}|\bmm{y},\bm{\theta})- \bmm{x} \|_2/ \|\bmm{x}\|_2 =0.179$ for the middle reconstruction, and $\|\mathbb{E}^\text{M1}(\bmm{x}|\bmm{y},\bm{\theta})- \bmm{x} \|_2/ \|\bmm{x}\|_2 =0.269$ for the rightmost one, while in Figure \ref{fig_krig_id_obs_sign05_2}, $\|\mathbb{E}^\text{M2}(\bmm{x}|\bmm{y},\bm{\theta})- \bmm{x} \|_2/ \|\bmm{x}\|_2 =0.168$ and $\|\mathbb{E}^\text{M1}(\bmm{x}|\bmm{y},\bm{\theta})- \bmm{x} \|_2/ \|\bmm{x}\|_2 =0.195$.

\subsubsection{AVA observations}
While the results using the identity observations are convincing in the extended models' favour, we also need to investigate the effects where the observation matrix is the seismic AVA model. In this case, the true fields are generated by Model 2, and the observations are linear combinations of the various fields at each space location followed by a convolution with a smooth wavelet and i.i.d noise. We use $\lambda^2=0.5$ and $\lambda^2=20$ in these examples. 

\begin{figure}
	\caption{Observations using identity observations (middle) and the seismic AVA model (bottom)} \label{fig_wa_observations_spde}
	\begin{center}
		\includegraphics[width=0.95\textwidth]{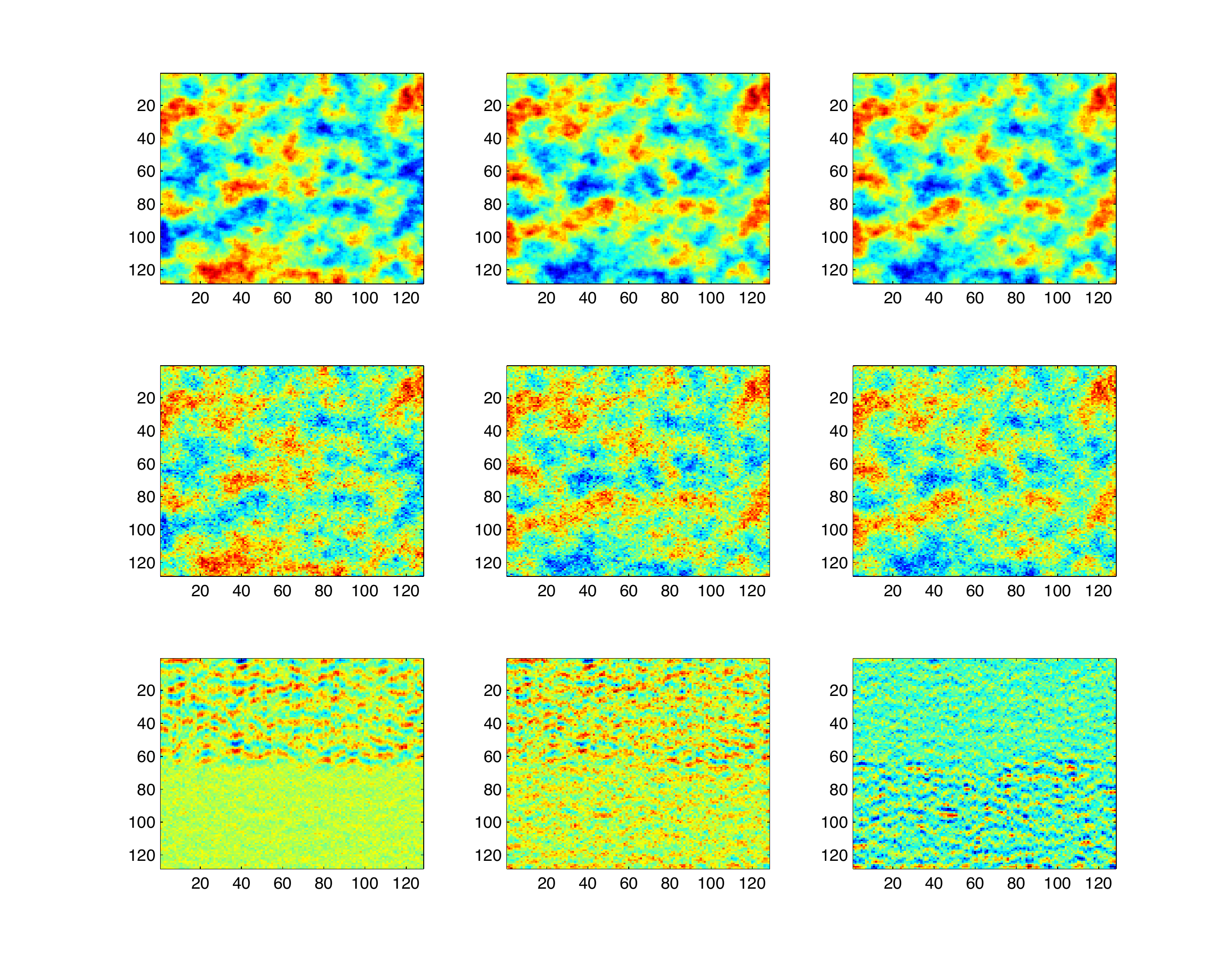}
	\end{center}
\end{figure}

In Figure \ref{fig_wa_observations_spde}, we can see the observations that are generated by this process. A key feature in the observations is that there occurs some cancellation, resulting from the fact that they are linear combinations of the underlying fields. This results in varying signal-to-noise ratios depending on the varying correlations.

\begin{figure}
	\caption{Reconstructed field using the AVA model with signal-to-noise ratio $1/0.5$. True parameters (left), kriging using Model 2 (centre), using Model 1 (right).} \label{fig_krig_WA_obs_sign05}
	\begin{center}
		\includegraphics[width=0.95\textwidth]{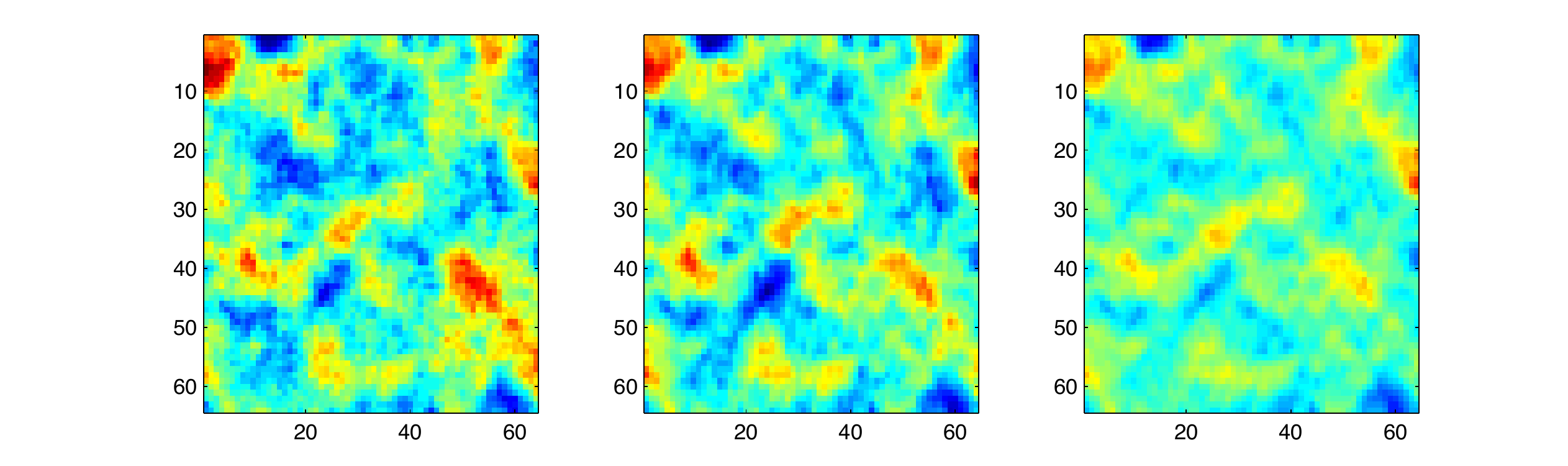}
	\end{center}
\end{figure}

\begin{figure}
	\caption{Reconstructed field using the AVA model with signal-to-noise ratio $1/20$. True parameters (left), kriging using Model 2 (centre), using Model 1 (right).} \label{fig_krig_WA_obs_sign20}
	\begin{center}
		\includegraphics[width=0.95\textwidth]{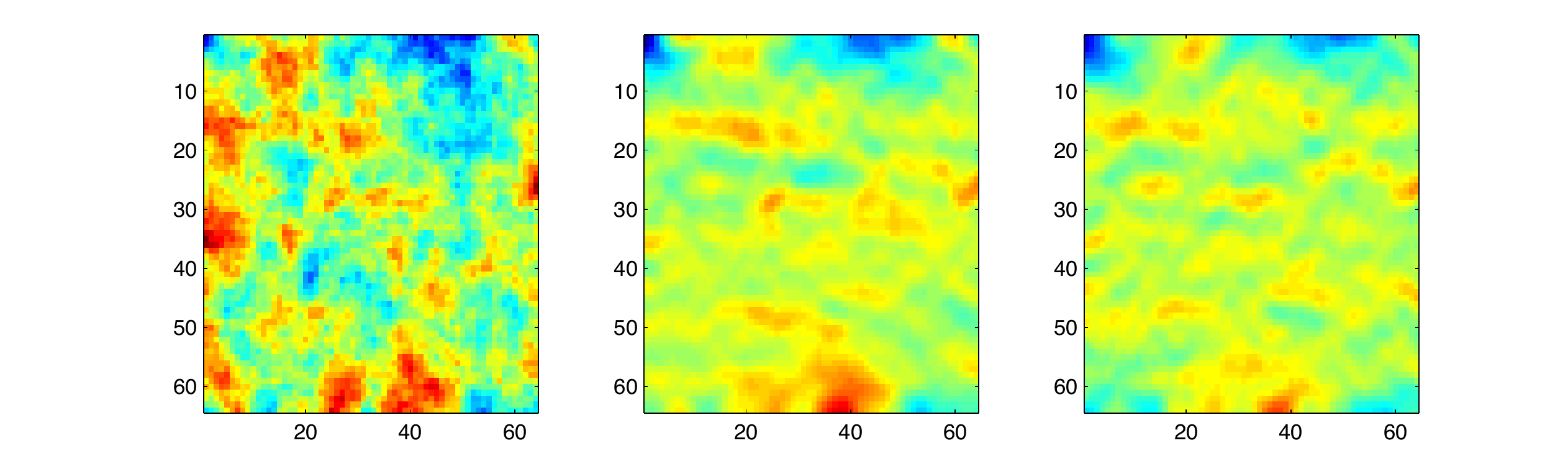}
	\end{center}
\end{figure}

Reconstructing the original multivariate field using the AVA observation scheme is more difficult than for using the identity observation matrix. The aforementioned cancellation effect is a major contributor to this. Additionally, there does not seem to be a straightforward way of interpreting the estimated correlation parameters coming from Model 1. In Figure \ref{fig_krig_WA_obs_sign05}, we illustrate the true parameters on the left, with reconstruction using the Model 2 the middle and the Model 1 on the right, using a signal-to-noise ratio of $1/0.5$. The effects we see are reminiscent of those using identity observations, but the smoothing effect is not present here. In this case $\|\mathbb{E}^\text{M2}(\bmm{x}|\bmm{y},\bm{\theta})- \bmm{x} \|_2/ \|\bmm{x}\|_2 =0.461$, while $\|\mathbb{E}^\text{M1}(\bmm{x}|\bmm{y},\bm{\theta})- \bmm{x} \|_2 / \|\bmm{x}\| =0.728$. Reconstruction using the same model, with a signal-to-noise ratio $1/20$ is depicted in Figure \ref{fig_krig_WA_obs_sign20}. No smoothing effect relative to Model 2 is observed here, but predictions are worse using Model 1, having $\|\mathbb{E}^\text{M2}(\bmm{x}|\bmm{y},\bm{\theta})- \bmm{x} \|_2/ \|\bmm{x}\|_2 =0.762$, while $\|\mathbb{E}^\text{M1}(\bmm{x}|\bmm{y},\bm{\theta})- \bmm{x} \|_2 / \|\bmm{x}\| =0.863$.

\subsubsection{Identity observations and non-stationarity}

\begin{figure}
	\caption{Kriging for identity observations with signal-to-noise ration $1/0.2$, field 1. True parameters (left), kriging using Model 3 (center), Model 1 (right)} \label{fig_krig_anis_nonstat_id_obs_sign02}
	\begin{center}
		\includegraphics[width=0.95\textwidth]{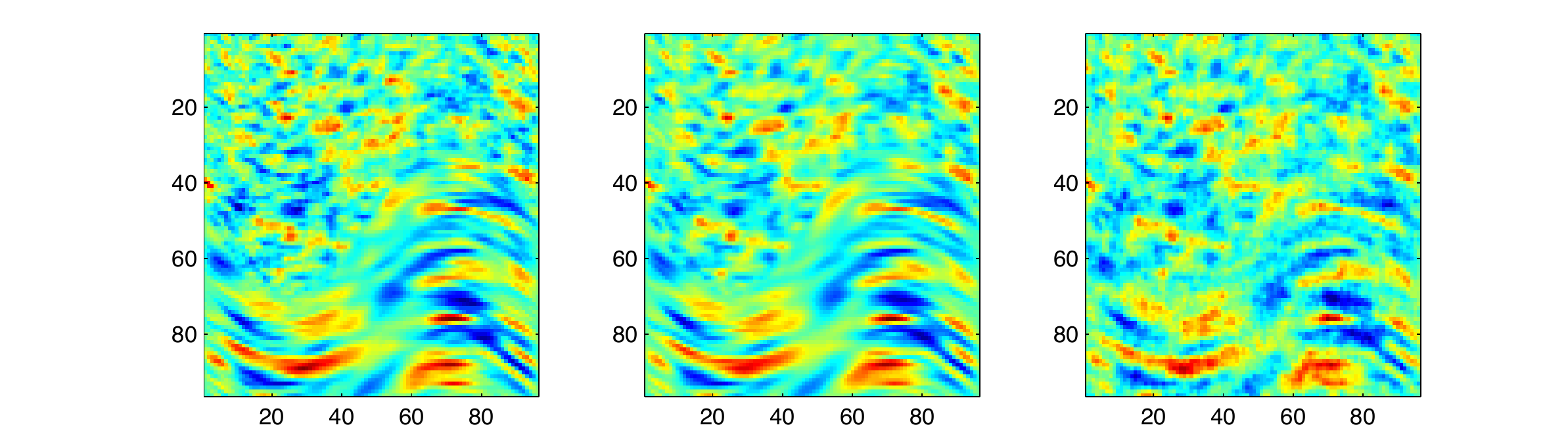}
	\end{center}
\end{figure}

Until this point, we have only studied the effects coming from changing correlations between interfaces. The model we have described is much richer than that, providing a flexible way of specifying anisotropy that moves along geometry of the subsurface. In this situation, we expect the results to be even more convincing, and we provide one example to cover this situation as well. In this case, we generate the true field by using Model 3, and we have identity observations followed by i.i.d. noise. The realisations of the true fields are then similar to the one in Figure \ref{fig_nonst_mvar_geos2}, and we estimate both the simple and complex model for thereafter giving a reconstruction of the latent field. In Figure \ref{fig_krig_anis_nonstat_id_obs_sign02}, we see the reconstructions using Model 3 (center) and Model 1 (right), and the most prominent effect we see is the smoothness differences in the bottom layer. Reconstruction using Model 1 is rugged and does not capture the anisotropy of the layer at all, contrasting the reconstruction using Model 3. On the top layer, on the other hand, the reconstructions are more comparable. The relative errors  are $\|\mathbb{E}^\text{M3}(\bmm{x}|\bmm{y},\bm{\theta})- \bmm{x} \|_2/ \|\bmm{x}\|_2 =0.280$ (center) and $\|\mathbb{E}^\text{M1}(\bmm{x}|\bmm{y},\bm{\theta})- \bmm{x} \|_2/ \|\bmm{x}\|_2 =0.382$ (right) for the reconstructions -- i.e. predictions are about 37\% better using the true model.

\subsubsection{Reconstruction -- final remarks}

It is also important to note that if we simulate from the simple model, the parameters here are recovered well by using the parametrisation in Section \ref{sec_param_q0}. This means that the correlations estimated by the simple model are close to the ones estimated by the more complex model. This, of course, adds to the usefulness of the model in situations where we do not know in advance that the correlation changes between interfaces. The uncertainty, however, is greater, leading to more disparate estimates of the correlations than when using the simple parametrisation.

\subsection{Estimating the blend range for fuzzy interfaces}

In this section, we will treat all parameter except the blend range as fixed. The model we will treat is one where the true interface is given as a sine function, and what we guess is a flat interface. This is exactly the model which is depicted in Figure \ref{fig_blend_range_geodesic}. In our example, we use Model 2 for constructing the true field, followed by identity observations and i.i.d. noise. 

\begin{figure}
	\caption{Sample from true model (top), sample from true minus guessed model (middle), sample from true model minus blend model with optimal range (bottom).} \label{fig_blend_sample_heur}
	\begin{center}
	\includegraphics[width=0.8\textwidth]{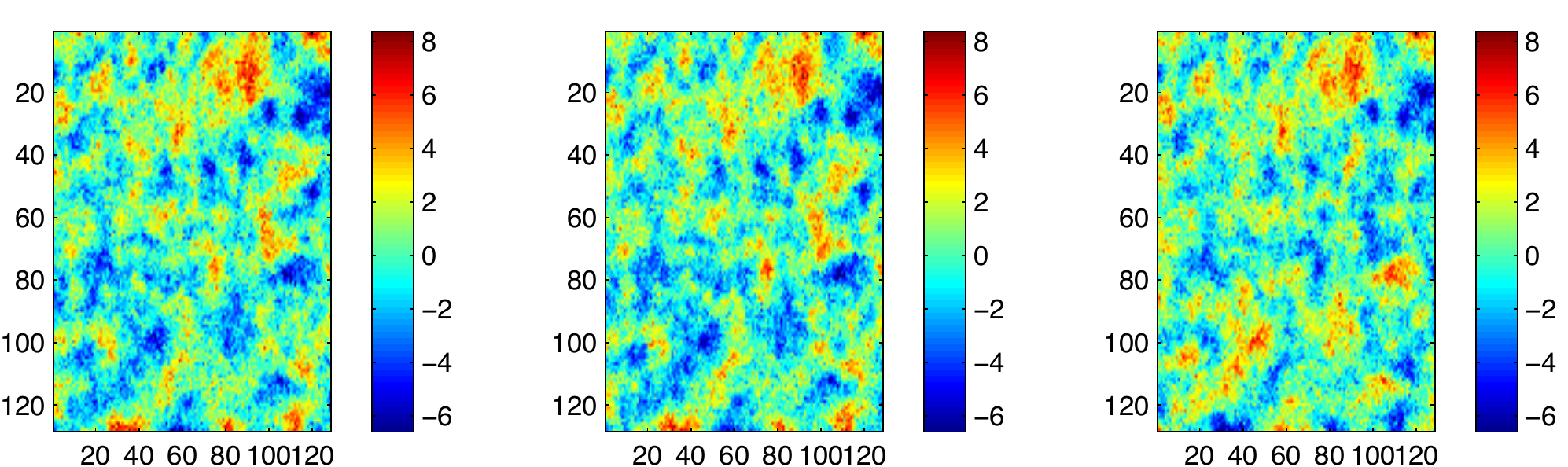}\\
	\includegraphics[width=0.8\textwidth]{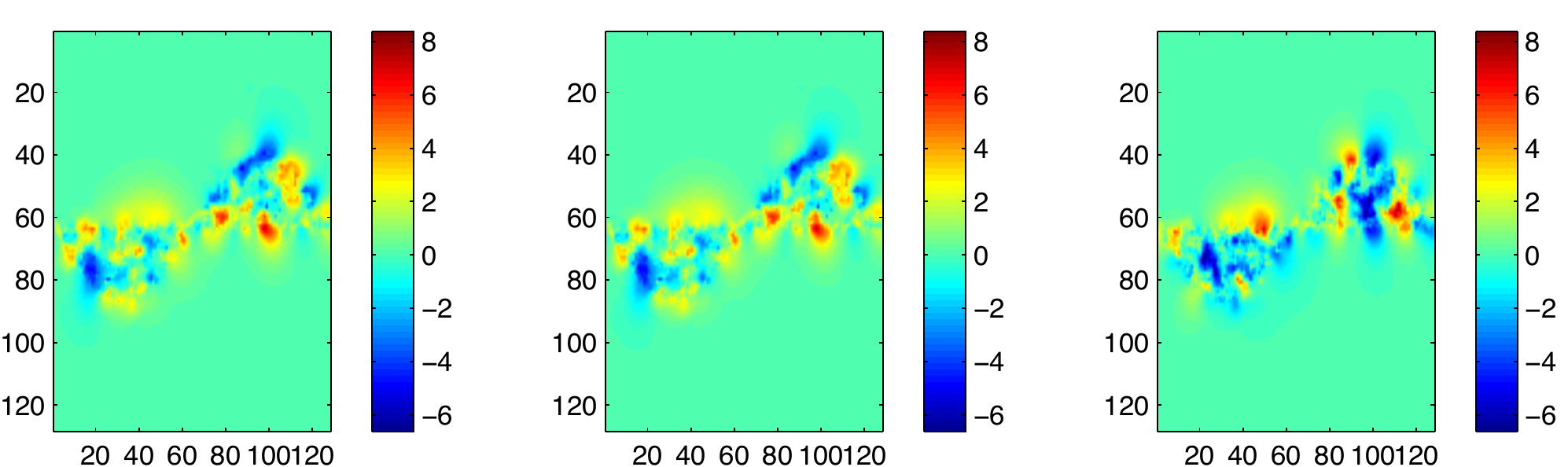}\\
	\includegraphics[width=0.8\textwidth]{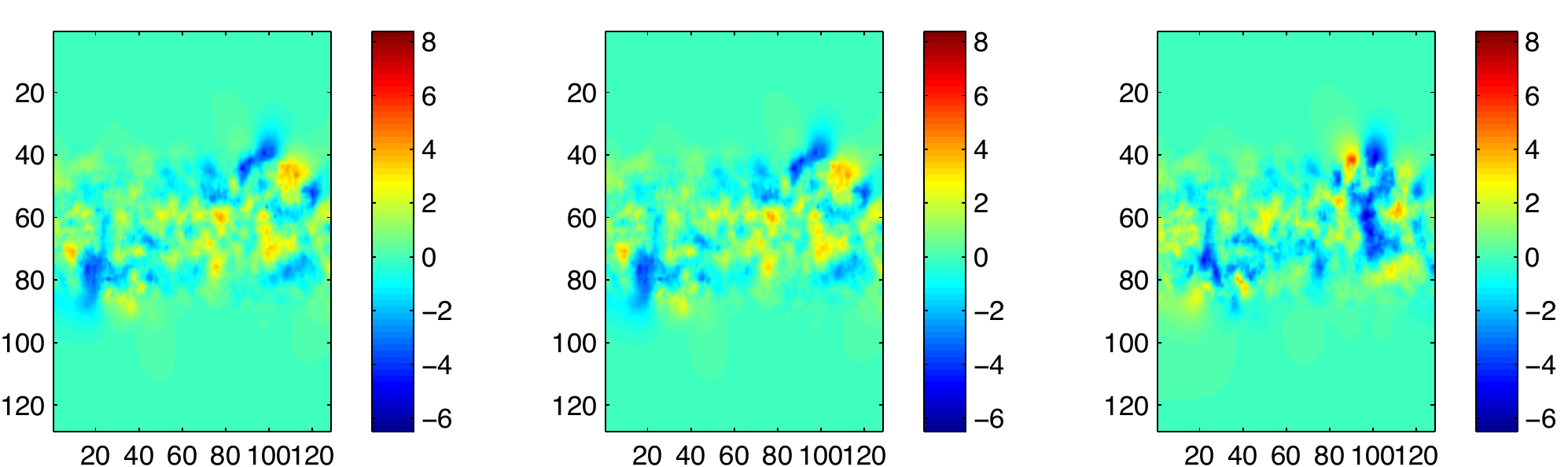}
	\end{center}
\end{figure}

Before actually doing maximum likelihood estimation, we visualise heuristically why it may make sense. In Figure \ref{fig_blend_sample_heur}, we see a sample of the true model at the top, the true sample minus the guessed model in the model, and the true sample minus the blend model with optimal range at the bottom. The norm of the bottom figure is less than that of the middle one.

\begin{figure}
	\caption{Estimated blend range using maximum likelihood for 200 samples. Sine-interface with full period (left), sine-interface with half period (right).} \label{fig_blendrange_ml}
	\begin{center}
		\includegraphics[width=0.45\textwidth]{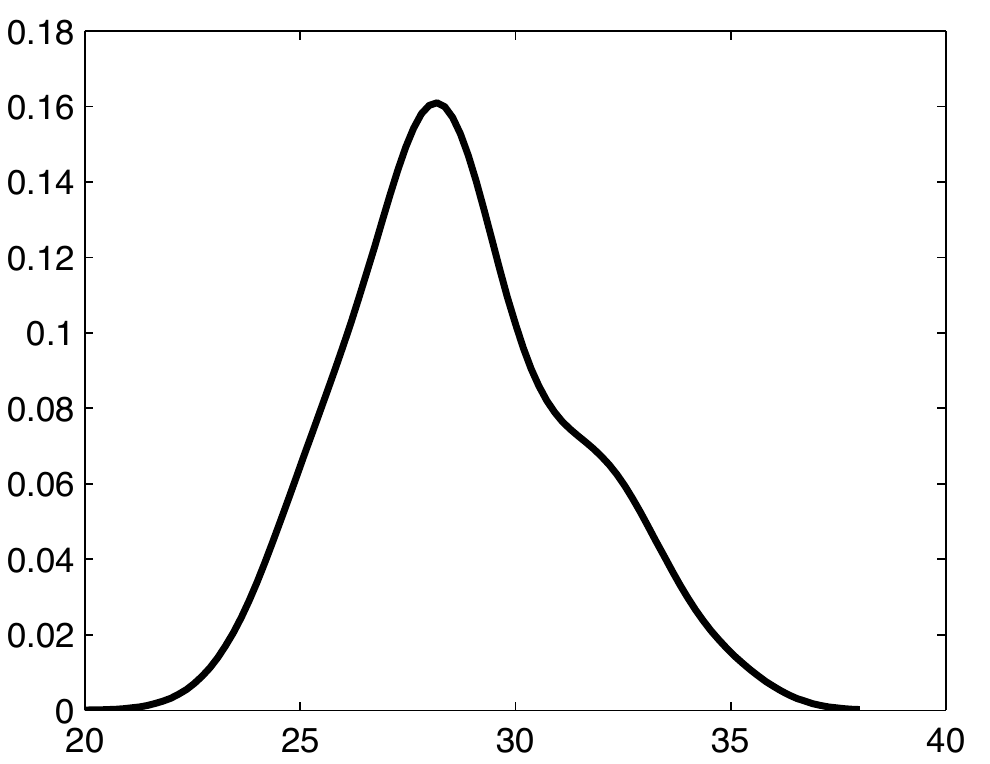} \includegraphics[width=0.45\textwidth]{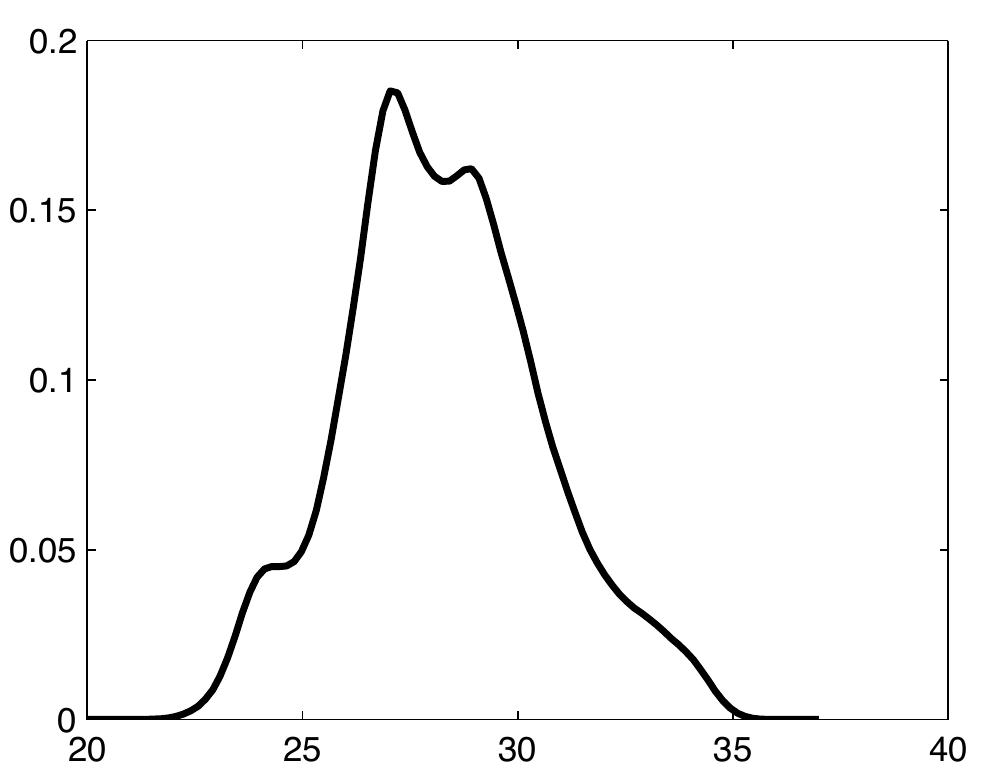}
	\end{center}
\end{figure}

Maximum likelihood estimates for the range is given in Figure \ref{fig_blendrange_ml}, where the left figure is the range estimates when the guessed interface is a line and the true line is a full-period sine with a maximum amplitude of $23$ and the right is a half-period sine with maximum amplitude $23$. These estimates are good in the sense that the range covers the true model as in Figure \ref{fig_blend_range_geodesic}.

\begin{figure}
	\caption{Performance gain using the blended interface over the guessed interface with sine-interface with full period (left) and sine-interface with half period (right)} \label{fig_blend_comp_perf}
	\begin{center}
		\includegraphics[width=0.45\textwidth]{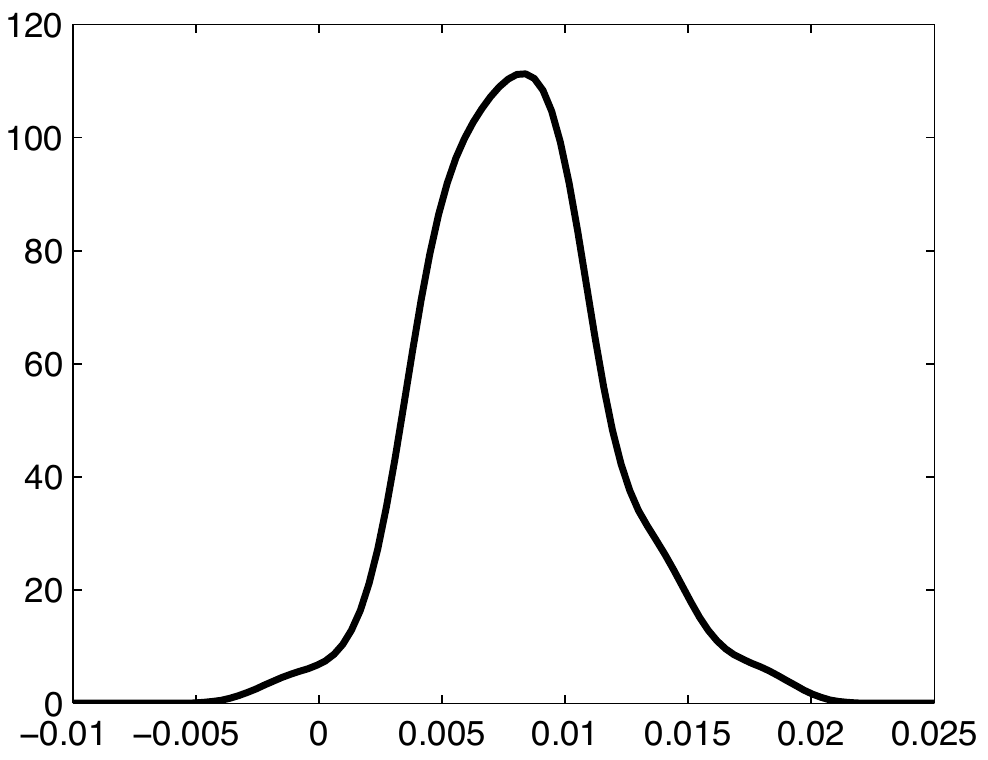} \includegraphics[width=0.45\textwidth]{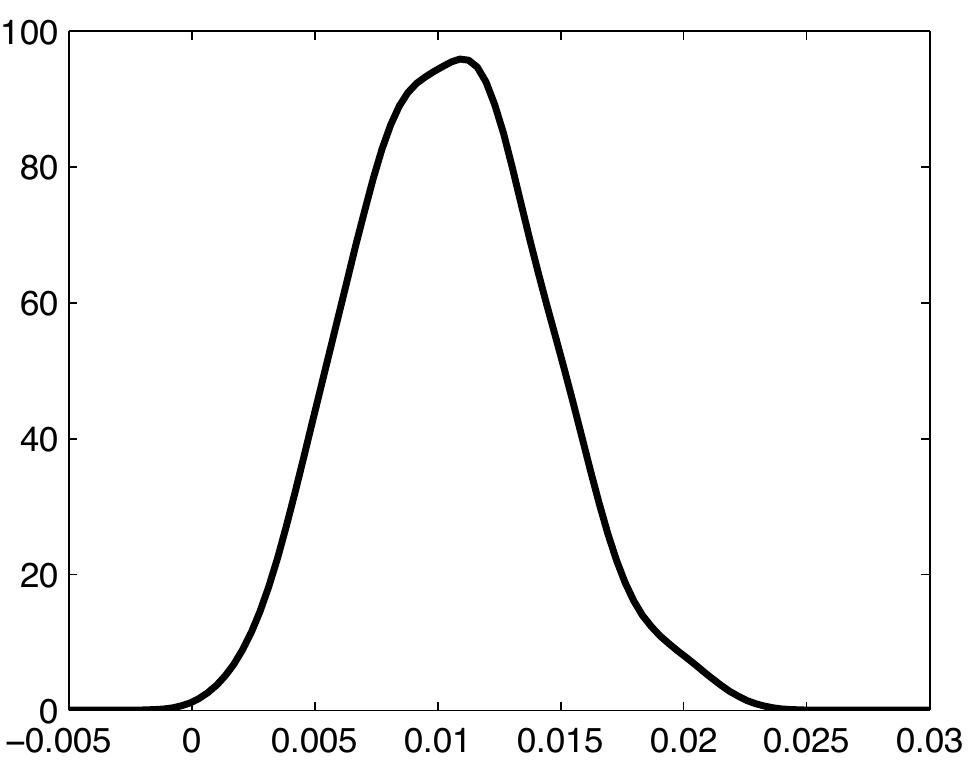}
	\end{center}
\end{figure}

A comparison of predictions using the guessed interface with no blend and the one with optimal blend is given in Figure \ref{fig_blend_comp_perf}. Here we see that the predictions using optimal blend range are only marginally better than using the interface with no blend for both interface structures. For the blend range, however, better prediction is not the goal. The goal here is to get an idea about how uncertain we are about the interface location, and better predictions comes as an additional boon, even if the improvement is marginal.

\section{Conclusions and future work}

In this text we have showed three things: First, how it is possible to incorporate information about the geometry of the problem flexibly. Secondly, how to facilitate changing covariances between elastic parameters depending on position. Lastly, we have introduced a novel way of specifying uncertainty related to the position of an interface using the concept of geodesic blending based on local correlation of the multivariate field. The first hinges on using SPDEs in order to specify local properties of the fields, and the second on how systems of SPDEs interrelate depending on position. The geodesic blending approach is based on the smooth manifold structure of the set of positive definite matrices. The ideas presented here are not limited to the relatively simple models described here -- rather, they may be used in any spatial inversion problem with a natural geometry where soft constraints based on expert opinion may be used.


\vspace{18pt}
\noindent \textbf{Address for corresponding author:} \\
Erlend Aune \\ Nedre M{\o}llenberggate 70B \\ 7043 Trondheim \\ E-mail: erlend.aune.1983@gmail.com

\newpage

\section*{Appendix: Finite difference disretization -- the gory details}
\addcontentsline{toc}{section}{Appendix: Finite difference disretization - the gory details}

This appendix is devoted to the finite difference scheme we used for discretizing the elliptic operator in \eqref{eq_matern_nonstat}. We employ a changed notation in this appendix for convenience, replacing $\bmm{H}$ with $\bmm{A}$, and we hope that it is transparent for readers. For a 2-dimensional field with $\alpha=1$, we have
\begin{align}
	&  \nabla \cdot \left ( \begin{array}{cc} a_{11}(x,y) & a_{12}(x,y) \\ a_{21}(x,y) & a_{22}(x,y) \end{array} \right ) \left ( \begin{array}{c} u_x(x,y) \\ u_y(x,y) \end{array} \right ) +\kappa(x,y) u(x,y) \notag  \\
	= &  \nabla \cdot \left( \begin{array}{c} a_{11}(x,y)u_x(x,y) + a_{1,2}(x,y)u_y(x,y) \notag \\ a_{21}(x,y)u_x(x,y) + a_{22}(x,y) u_y(x,y) \end{array} \right ) + \kappa(x,y) u(x,y) \notag \\
	= &  \partial_x(a_{11}u_x +a_{12}u_y) + \partial_y(a_{21} u_x + a_{22} u_y) +\kappa u \notag  \\
	= & a_{11}^x u_x + a_{11} u_{xx} + a_{12}^x u_y + a_{12} u_{yx} +a_{21}^y u_x + a_{21} u_{xy} + a_{22}^y u_{y} + a_{22} u_{yy} \notag \\
	= & \textnormal{diag}(A) \nabla \cdot \nabla u + (a_{12}+ a_{21}) u_{xy} +a^x_{11} u_x +a^x_{12} u_y + a_{22}^y u_y + a_{21}^y u_x  \label{eq_expl_2d}
\end{align}
where $a_{ij}^v, v=x,y$ denotes differentation wrt. $x$ or $y$ of the $i,j$ element of $\bmm{A}$, depending implicitly on the position. To discretize \eqref{eq_expl_2d}, we employ a finite difference scheme. We define the following finite difference operators
\begin{align}
	\delta_{x} u &= \frac{1}{h} (u_{i+1}^j - u_i^j) \notag \\
	\delta_{\hat{x}} &= \frac{1}{h} (u_i^j-u_{i-1}^j), \notag
\end{align}
where $i,j$ are positions on the grid, with $i$ denoting the $x$-direction and $j$ denoting the $y$-direction. Now, we define the following operators
\begin{align}
	\Lambda_{xx}u&=\delta_x \left ( \alpha_{11} \delta_{\hat{x}} u \right) =\delta_x \left ( \frac{1}{h} \alpha_{11} \left(u_i^j-u_{i-1}^j \right) \right ) \notag \\
	&=\frac{1}{h^2}\left (  \alpha_{11}^{i+1,j} \left( u_{i+1}^j-u_i^j \right ) - \alpha_{11}^{i,j} \left ( u_i^j - u_{i-1}^j \right ) \right ),
\end{align}
where
\begin{align}
	\alpha_{11}^{i,j} &= \frac{1}{2} \left( a_{11}^{i,j} +a_{11}^{i-1,j} \right) \notag \\
	\alpha_{22}^{i,j} &= \frac{1}{2} \left( a_{22}^{i,j} +a_{22}^{i,j-1} \right). \notag
\end{align}
A equivalent expression holds for $\Lambda_{yy} u$. We define $\alpha_{kk}^{1,1}=a_{11}^{1,1}, k=1,2$. For the mixed operators we have
\begin{align}
	\Lambda_{xy}^+ u & = \frac{1}{2} \left (  \delta_x \left( a_{12} \delta_y u \right) + \delta_{\hat{x}} \left ( a_{12} \delta_{\hat{y}} u \right) \right)
\end{align}
and we have 
\begin{align}
	\delta_x \left(a_{12} \delta_y u \right) & = \frac{1}{h} \delta_x \left(a_{12} u_i^{j+1}-u_i^j \right) \notag \\
	& = \frac{1}{h^2} \left( a_{12}^{i+1,j} (u_{i+1}^{j+1}-u_{i+1}^j ) - a_{12}^{i,j}(u_i^{j+1} - u_i^j)  \right ) \notag \\
	\delta_{\hat{x}} \left( a_{12} \delta_{\hat{y}} u \right ) &=\frac{1}{h} \delta_{\hat{x}} \left( a_{12} (u_i^j - u_i^{j-1}) \right) \notag \\
	& = \frac{1}{h^2} \left( a_{12}^{i,j}(u_i^j - u_i^{j-1}) - a_{12}^{i-1,j}(u_{i-1}^j - u_{i-1}^{j-1}) \right ). \notag
\end{align}
Hence
\begin{align}
	\Lambda_{xy}^+ u = & \frac{1}{2 h^2}\Big( \Big(a_{12}^{i+1,j} ( u_{i+1}^{j+1} - u_{i+1}^j) - a_{12}^{i,j} ( u_i^{j+1} - u_i^j) \Big)  \\
	& + \Big(a_{12}^{i,j}(u_i^j -u_i^{j-1}) - a_{12}^{i-1,j}(u_{i-1}^j-u_{i-1}^{j-1} ) \Big) \Big) \notag
\end{align}
For $\Lambda_{yx}^+$ we reverse the order of the difference operators:
\begin{align}
	\Lambda_{yx}^+ u & = \frac{1}{2}(\delta_{y}(a_{12}\delta_x u) + \delta_{\hat{y}} (a_{12} \delta_{\hat{x}})) \notag \\
	&=\frac{1}{2h^2} \Big( \Big(a_{12}^{i,j+1}(u_{i+1}^{j+1} -u_i^{j+1}) -a_{12}^{i,j}(u_{i+1}^j - u_i^j) \Big) \notag \\
	& \quad + \Big(a_{12}^{i,j}(u_i^j-u_{i-1}^j) - a_{12}^{i,j-1}(u_{i}^{j-1} - u_{i-1}^{j-1}  \Big) \Big)  \notag
\end{align}
And the complete discretisation is
\begin{align}
	(\Lambda_{xx} + \Lambda_{xy}^+ + \Lambda_{yx}^+ + \Lambda_{yy})u =f(u,W)
\end{align}
In \cite{monotone_diff_2d}, it is proved that this scheme is convergent. If we assume that $\bmm{A}$ does not vary in space, we can simplify the scheme;
\begin{align}
	\widehat{\Lambda_{xx}} u & = \frac{1}{h^2} (a_{11}(u_{i+1}^j -u_i^j) - a_{11}(u_i^j-u_{i-1}^j))  \notag \\
	& = \frac{1}{h^2} a_{11}(u_{i+1}^j -2u_i^j +u_{i-1}^j) \notag \\
	\widehat{\Lambda_{yy}}u & = \frac{1}{h^2} a_{22}(u_{i}^{j+1} -2u_i^j +u_{i}^{j-1}) \notag \\
	\widehat{\Lambda_{xy}^+} u & = \frac{1}{2 h^2} \Big( \Big(a_{12} ( u_{i+1}^{j+1} - u_{i+1}^j) - a_{12} ( u_i^{j+1} - u_i^j) \Big) \notag \\
	& \quad + \Big(a_{12}(u_i^j -u_i^{j-1}) - a_{12}(u_{i-1}^j-u_{i-1}^{j-1}) \Big) \Big) \notag \\
		& = \frac{a_{12}}{2 h^2} \left(2u_i^j +u_{i+1}^{j+1}+u_{i-1}^{j-1} - u_{i+1}^j - u_i^{j+1} -u_i^{j-1} - u_{i-1}^j \right) \notag \\
		\widehat{\Lambda_{yx}^+} u & = \frac{1}{2h^2} \Big( \Big(a_{12}(u_{i+1}^{j+1} -u_i^{j+1}) -a_{12}(u_{i+1}^j - u_i^j) \Big) \notag \\
		&\quad  + \Big(a_{12}(u_i^j-u_{i-1}^j) - a_{12}(u_{i}^{j-1} - u_{i-1}^{j-1}  \Big) \Big) \notag \\
			&= \frac{a_{12}}{2 h^2} \left( 2 u_i^j +u_{i+1}^{j+1} + u_{i-1}^{j-1} - u_{i+1}^j - u_i^{j+1} -u_i^{j-1} -u_{i-1}^j     \right) \notag \\
		\left(\widehat{\Lambda_{xy}^+}+ \widehat{\Lambda_{yx}^+} \right)u & = \frac{a_{12}}{h^2} \left( 2 u_i^j +u_{i+1}^{j+1} + u_{i-1}^{j-1} - u_{i+1}^j - u_i^{j+1} -u_i^{j-1} -u_{i-1}^j     \right) \notag
\end{align}
This corresponds to the following stencil
\begin{align}
	S =-\frac{1}{h^2}\left( \begin{array}{ccc} a_{12}	&	-a_{22}-a_{12}	&	0 \\ -a_{11} - a_{12} & 2(a_{11}+a_{22}+a_{12}) & -a_{11} - a_{12} \\ 0 & -a_{22}-a_{12} & a_{12} \end{array} \right)
\end{align}

\newpage

\bibliography{/home/anne/erlenda/Dropbox/ntnu/tex/bible/bible}
\bibliographystyle{apa}

\end{document}